\documentclass[a4paper,aps,pre,superscriptaddress,floatfix,nofootinbib,longbibliography,notitlepage,onecolumn]{revtex4-1}
\usepackage{graphicx, subfigure}
\usepackage{amssymb, amsmath,amssymb,amsfonts}
\usepackage{amsthm,mathrsfs,amsopn}
\usepackage{dcolumn}
\usepackage{bm}
\usepackage{xcolor}
\usepackage[utf8]{inputenc}
\usepackage{mathtools}
\usepackage[english]{babel}

\usepackage{graphicx} 

\theoremstyle{plain} 

\newtheorem{remark}{Remark}

\begin{document}

\title{Synchronization of nonlinearly coupled Stuart-Landau oscillators on networks}

\author{Wilfried Segnou}
\affiliation{Department of mathematics and naXys, Namur Institute for Complex Systems, University of Namur, Belgium}

\author{Riccardo Muolo}
\affiliation{Department of Systems and Control Engineering, Institute of Science Tokyo (former Tokyo Tech), Japan}

\author{Marie Dorchain}
\affiliation{Department of mathematics and naXys, Namur Institute for Complex Systems, University of Namur, Belgium}

\author{Hiroya Nakao}
\affiliation{Department of Systems and Control Engineering, Institute of Science Tokyo (former Tokyo Tech), Japan}
\affiliation{International Research Frontiers Initiative, Institute of Science Tokyo (former Tokyo Tech), Japan}

\author{Timoteo Carletti}
\affiliation{Department of mathematics and naXys, Namur Institute for Complex Systems, University of Namur, Belgium}

\begin{abstract}
The dynamics of coupled Stuart–Landau oscillators play a central role in the study of synchronization phenomena. Previous works have focused on linearly coupled oscillators in different configurations, such as all-to-all or generic complex networks, allowing for both reciprocal or non-reciprocal links. The emergence of synchronization can be deduced by proving the linear stability of the limit cycle solution for the Stuart-Landau model; the linear coupling assumption allows for a complete analytical treatment of the problem, mostly because the linearized system turns out to be autonomous. In this work, we analyze Stuart–Landau oscillators coupled through nonlinear functions on both undirected and directed networks; synchronization now depends on the study of a non-autonomous linear system and thus novel tools are required to tackle the problem. We provide a complete analytical description of the system for some choices of the nonlinear coupling, e.g., in the resonant case. Otherwise, we develop a semi-analytical framework based on Jacobi-Anger expansion and Floquet theory, which allows us to derive precise conditions for the emergence of complete synchronization. The obtained results extend the classical theory of coupled oscillators and pave the way for future studies of nonlinear interactions in networks of oscillators and beyond.
\end{abstract}

\maketitle

\section{Introduction}

Synchronization is one of the most astonishing and widespread phenomenon of self-organization in complex systems. Being observed in many natural and engineered systems, it finds applications in various fields, from neuroscience to mechanics, to superconductivity and power grids~\cite{pikovsky2001synchronization,Strogatzbooksync}. 

Synchronization emerges from the joint action of coupled self-sustained oscillators, being the latter regular ones, i.e., periodic, or chaotic systems~\cite{pikovsky2001synchronization}. Mathematically, a stable periodic solution can be represented by a limit cycle, i.e., a closed isolated attractive trajectory~\cite{pikovsky2001synchronization,strogatz2018nonlinear}. If it emerges through a supercritical Hopf-Andronov bifurcation~\cite{strogatz2018nonlinear}, by exploiting the center-manifold reduction, one can prove that the Stuart-Landau (SL) model~\cite{garcia2012complex,nakao2014complex} results to be a normal form allowing to describe the behavior of a generic oscillator in this framework~\cite{kuramoto2019concept}. Moreover, by performing  the phase reduction~\cite{nakao16,monga2019phase} for a system of globally coupled SL oscillators, Yoshiki Kuramoto obtained his celebrated model~\cite{kuramoto1975} in 1975, which is nowadays one of the most studied models in nonlinear science~\cite{acebron2005kuramoto}. These two examples provide a strong support for the universality and importance of the dynamics of coupled SL oscillators, and the reason why it is nowadays a paradigm in the study of synchronization dynamics.

Note that differential equations describing coupled Stuart-Landau oscillators are also called Complex Ginzburg-Landau Equation (CGLE), in particular in the continuous setting, which derives from center manifold reduction of continuous reaction-diffusion equations describing pattern formation and oscillations in chemical systems near the supercritical Hopf bifurcation~\cite{kuramoto1975formation}. The spatiotemporal chaotic state of CGLE, which occurs via Benjamin-Feir instability of the uniformly synchronized state, has been extensively studied~\cite{kuramoto1984chemical,aranson2002world}. Other types of interactions between the oscillators in discrete settings have also been considered. First, globally coupled SL oscillators, namely all-to-all and mean-field interactions, were introduced and unveiled a rich behavior, from cluster synchronization to collective chaos ~\cite{hakim1992dynamics,nakagawa1993collective,nakagawa1994collective,chabanol1997collective,banaji1999towards,daido2006diffusion,daido2007aging,ku2015dynamical,kemeth2019cluster,zajdela2025phase,thome2025hierarchical}. Then, nonlocally coupled SL oscillators were introduced~\cite{kuramoto1995scaling,kuramoto1996origin,kuramoto1997power,kuramoto1998multiaffine,nakao1999anomalous,garcia2008coherent,lee2022nontrivial} and anomalous spatiotemporal chaos with fractal structures and chimera states were found.

After the birth of network science in the late 1990s~\cite{newmanbook,Latorabook}, synchronization became a hot topic in the study of dynamics on networks~\cite{arenas2008synchronization,boccaletti2018synchronization} and the Stuart-Landau system acquired a central role~\cite{nakao2014complex,PhysRevE.93.032317,dipatti2} also in this framework. Different settings have been analyzed, e.g., chaos has been found by coupling SL oscillators with scale free networks~\cite{nakao2009diffusion}; chimera states emerge once SL oscillators are coupled via nonlocal rings~\cite{ZKS2014} or by using two populations of SL oscillators coupled with an all-to-all graph~\cite{Laing2010}. Furthermore, it was shown that directed networks enhance the emergence of the so-called Benjamin-Feir instability~\cite{di2017benjamin} and that synchronization can be enhanced through time-varying networks~\cite{pereti2020stabilizing}, to name a few. It is worth mentioning also the emergence of remote synchronization, first observed in the framework of phase oscillators~\cite{okuda1991mutual}, but then also fully described in systems of Stuart-Landau oscillators~\cite{bergner2012remote}. It is important to note that in all the above works the coupling between the Stuart-Landau oscillators is assumed to be linear. Nonlinear global coupling in the case of continuous support, has been considered in~\cite{schmidt2014coexistence}, where it was shown, theoretically and experimentally, the emergence of chimera states.

In this work, we focus on  \emph{complete synchronization} of identical Stuart-Landau oscillators nonlinearly coupled via a complex network~\footnote{Note that, in the context of higher-order networks and topological signals, complete synchronization is sometimes called \emph{global synchronization}~\cite{carletti2023global,wang2024global,carletti2025global}, which might be confusing in this context. In fact, the term \emph{global} could be misleading, by suggesting that the system synchronizes for any initial condition. In fact, our analysis is local and applies only for initial conditions close to the synchronous solution.}. The present study is based on the Master Stability Function~\cite{fujisaka1983stability,pecora1998master}; this method is rooted on a linear stability analysis technique consisting of perturbing the system about the synchronous solution, i.e., the stable limit cycle of each isolated SL oscillator, and then to reduce the dimension of the resulting linearized system by exploiting the eigenbasis of a suitable operator encoding for the coupling, i.e., the Laplace matrix. From the stability property of the resulting low dimensional linearized system, one can infer the emergence of synchronization. Under the assumption of nonlinear coupling, this recipe can be applied, however the eventual linearized system results to be non-autonomous and thus its analysis turns out to be quite involved. We found that, for certain nonlinear coupling functions, e.g., the resonant power law (see hereby), despite the nonlinearity, we are able to carry out a full analytical treatment of the problem and obtain conditions for complete synchronization or, alternatively, for an instability. Note that the nonlinear coupling discussed in~\cite{schmidt2014coexistence} is reminiscent of this type. On the other hand, general nonlinear couplings do not allow for an analytical treatment. Nonetheless, in the case of non-resonant power laws, we developed a semi-analytical framework that combines the Jacobi–Anger expansion with Floquet theory, by allowing us to derive approximate conditions under which complete synchronization is obtained. The proposed theory is complemented by dedicated numerical simulations of SL oscillators nonlinearly coupled by using complex networks with diverse topologies, e.g., scale-free, Watts-Strogatz or random graphs.

\section{The model}
\label{sec:model}

Let us consider $N$ identical \textit{Stuart-Landau} (SL) oscillators and assume they are nonlinearly coupled together by pairwise connections. More precisely, the time evolution of the state variable of the $j$-th oscillator, $W_j(t) \in \mathbb{C}$, is given by
\begin{equation}
\label{eq:SLmodelGraph}
    \frac{dW_{j}}{dt} = \sigma W_{j} - \beta |W_{j}|^{2}W_{j} + \mu \sum_{k}{\Delta}_{jk}f\left(W_{k},\bar{W}_{k}\right)\, ,
\end{equation}
where $f(W,\bar{W})$ is a nonlinear function depending on the variable $W$ and its complex conjugate $\bar{W}$, $\sigma = \sigma_{\Re} + i\sigma_{\Im}$ and $\beta = \beta_{\Re} + i\beta_{\Im}$ are the complex parameters of the Stuart-Landau model, and $\mu = \mu_{\Re} + i\mu_{\Im}$ is the complex coupling strength. The network structure is encoded by the Laplace matrix, $\pmb{\Delta}=\mathbf{A}-\mathbf{D}$, where $\mathbf{A}$ is the adjacency matrix, ${A}_{jk} = 1$ if there is a directed edge from node $k$ to node $j$, ${A}_{jk} = 0$ otherwise, and $\mathbf{D}$ is the diagonal degree matrix, $D_{kk}=\sum_j A_{jk}$.

For the sake of pedagogy, let us assume to consider the Taylor expansion of $f$ and to retain only the lower order terms. Namely, we will consider 
\begin{equation}
\label{eq:SLmodelGraph2}
    \frac{dW_{j}}{dt} = \sigma W_{j} - \beta |W_{j}|^{2}W_{j} + \mu \sum_{k}{\Delta}_{jk}W_{k}^a\bar{W}_{k}^b\, ,
\end{equation}
for some integers $a$ and $b$, which control the nonlinearity of the coupling term~\cite{pietras2019network}. Let us observe that the possible term of $f$ associated to $a=b=0$ cancels out because of the property of the Laplace matrix $\sum_j \Delta_{jk}=0$ for all $k$. The case $a=1$ and $b=0$, i.e., corresponding to linear coupling, has been largely analyzed in the literature, as discussed in the Introduction. The choice $a=2$ and $b=1$ is reminiscent of the nonlinear coupling discussed in~\cite{schmidt2014coexistence}. Let us stress that the theory hereby developed applies to the more general case of real $a$ and $b$.

In the \textit{supercritical regime}, $\sigma_\Re >0$ and $\beta_\Re>0$, each isolated SL oscillator admits a stable limit cycle solution~\cite{nakao2014complex,kuramoto1984chemical} of the form $W_{\textrm{LC}}(t) = |W_{\textrm{LC}}| \exp(i \omega t)$, with amplitude $|W_{\textrm{LC}}| = \sqrt{\dfrac{\sigma_{\Re}}{\beta_{\Re}}}$ and frequency $\omega = \sigma_{\Im} - \beta_{\Im} \dfrac{\sigma_{\Re}}{\beta_{\Re}}$. Complete synchronization of the system of coupled SL oscillators requires $W_j(t)=W_{\textrm{LC}}(t)$ for all $j=1,\dots,N$ to be a stable solution of~\eqref{eq:SLmodelGraph2}. The latter is known as the synchronization manifold. By using again the property of the Laplace matrix, $\sum_j \Delta_{jk}=0$, it is trivial to show that indeed, $W_j(t)=W_{\textrm{LC}}(t)$ for all $j=1,\dots,N$, solves the system~\eqref{eq:SLmodelGraph2}. The proof of its stability under suitable conditions will be the aim of the next Section.

\section{Linear stability analysis}
\label{sec:linstabanal}

To prove the stability of the synchronization manifold, we will resort to a linear stability analysis close to the limit cycle solution. Let us thus consider a heterogeneous perturbation about the solution $W_{\textrm{LC}}(t)$:
\begin{equation}
\label{eq:perturb}
W_j(t) = W_{\textrm{LC}}(t)(1 + \rho_j(t))e^{i\theta_j(t)} \quad \forall j=1,\dots, N\, .
\end{equation}
$\rho_j$ and $\theta_j$ are small perturbations whose time evolution can be obtained by substituting the above expression into~\eqref{eq:SLmodelGraph} and retrieving only first order terms
\begin{equation} \label{eq:SLLinearmodelGraph2}
    \left\{
    \begin{aligned}
        \frac{d{\rho}_j}{dt} &= -2\sigma_{\Re}\rho_j 
        + |W_{LC}|^{a+b-1} \sum_k {\Delta}_{jk} 
        \left[(a+b)T_1(t) \rho_k - (a-b)T_2(t) \theta_k \right] \\
        \frac{d{\theta}_j}{dt} &= -2\beta_{\Im}\dfrac{\sigma_{\Re}}{\beta_{\Re}}\rho_j 
        + |W_{LC}|^{a+b-1} \sum_k {\Delta}_{jk} 
        \left[(a+b)T_2(t) \rho_k + (a-b)T_1(t) \theta_k \right]\, ,
    \end{aligned}
    \right. 
\end{equation}

where
\begin{equation}
\label{eq:T1T2}
\begin{aligned}
T_1(t) &= \mu_{\Re}\cos[(a-b-1)\omega t] - \mu_{\Im}\sin[(a-b-1)\omega t],\\
T_2(t) &= \mu_{\Im}\cos[(a-b-1)\omega t] + \mu_{\Re}\sin[(a-b-1)\omega t].
\end{aligned}
\end{equation}

Let us assume the Laplacian matrix to be diagonalizable, i.e., for each $\alpha=1,\dots N$, there exists an eigenvector $\Psi^{(\alpha)}$ with eigenvalue $\Lambda^{(\alpha)}$ such that $\pmb{\Delta} \Psi^{(\alpha)} = \Lambda^{(\alpha)} \Psi^{(\alpha)}$. By projecting the perturbations $\rho_j$ and $\theta_j$ onto the eigenbasis of the Laplacian matrix $\pmb{\Delta}$,
\begin{equation}
\label{eq:projeigenb}
\rho_j = \displaystyle\sum_{\alpha} \hat{\rho}_\alpha \Psi_j^{(\alpha)} \text{ and }
\theta_j = \displaystyle\sum_{\alpha} \hat{\theta}_\alpha \Psi_j^{(\alpha)}\, ,
\end{equation}
we can rewrite the linearized system~\eqref{eq:SLLinearmodelGraph2} in terms of $\hat{\rho}_\alpha$ and $\hat{\theta}_\alpha$, namely,
\begin{equation}
\label{eq:SLLinearmodelGraphTimDep}
\begin{aligned}
\frac{d}{dt}
\begin{pmatrix}
    \hat{\rho}_{\alpha} \\
    \hat{\theta}_{\alpha}
\end{pmatrix}
&=
\Bigg[
\begin{pmatrix}
    -2\sigma_{\Re} & 0 \\
    -2\beta_{\Im}\dfrac{\sigma_{\Re}}{\beta_{\Re}} & 0
\end{pmatrix}
+ |W_{LC}|^{a+b-1}\Lambda^{(\alpha)}
\begin{pmatrix}
    (a+b)T_{1}(t) & -(a-b)T_{2}(t) \\
    (a+b)T_{2}(t) & (a-b)T_{1}(t)
\end{pmatrix}
\Bigg]
\begin{pmatrix}
    \hat{\rho}_{\alpha} \\
    \hat{\theta}_{\alpha}
\end{pmatrix}
\\
&=: \mathbf{J}_{\alpha}(t)
\begin{pmatrix}
    \hat{\rho}_{\alpha} \\
    \hat{\theta}_{\alpha}
\end{pmatrix}\, .
\end{aligned}
\end{equation}

We can observe that the stability problem reduces to the study of the $1$-parameter family of $2\times2$ linear systems, each one depending on the Laplace eigenvalue $\Lambda^{(\alpha)}$. If $a\neq b+1$ the matrix $\mathbf{J}_{\alpha}(t)$ is $T$-periodic, $T=2\pi/[(a-b-1)\omega]$. On the other hand, if the {\em resonance condition} $a= b+1$ is satisfied, the matrix is autonomous, because $T_1 = \mu_{\Re}$ and $ T_2 = \mu_{\Im}$. Let us note that $\Lambda^{(1)}=0$, and thus $\mathbf{J}_{1}(t)$ is always time independent and coincides with the Jacobian of the isolated SL systems, which results to be marginally stable being its eigenvalues $0$ and $-2\sigma_\Re$. In conclusion, if, for all $\alpha =2,\dots, N$, the matrix $\mathbf{J}_{\alpha}(t)$ admits a negative Lyapunov exponent, then the system~\eqref{eq:SLmodelGraph2} admits complete synchronization. On the contrary, the solution $W_j(t)=W_{\mathrm{LC}}(t)$ is unstable and the system cannot synchronize.

To continue the stability analysis, we will consider separately the case $a=b+1$ and $a\neq b+1$. In the first one, the knowledge of the spectrum of $\mathbf{J}_{\alpha}$ will be enough to conclude; in the second one, we will resort to Floquet theory to analyze the stability of the synchronization manifold.

\subsection{The resonant case: Autonomous linear equation}
\label{ssec:autonom}

Let us thus assume $a=b+1$. A straightforward computation returns
\begin{equation}
\label{eq:Jalphaaut}
\mathbf{J}_{\alpha} =
\begin{pmatrix}
-2\sigma_{\Re} + (2a - 1)|W_{LC}|^{2(a - 1)}\mu_{\Re}\Lambda^{(\alpha)} & 
-|W_{LC}|^{2(a - 1)}\mu_{\Im}\Lambda^{(\alpha)} \\
-2\beta_{\Im}\dfrac{\sigma_{\Re}}{\beta_{\Re}} + (2a - 1)|W_{LC}|^{2(a - 1)}\mu_{\Im}\Lambda^{(\alpha)} & 
|W_{LC}|^{2(a - 1)}\mu_{\Re}\Lambda^{(\alpha)}
\end{pmatrix}=:\mathbf{J}_{0}+\Lambda^{(\alpha)}\mathbf{J}_{h}\, ,
\end{equation}
where we introduced
\begin{equation}
\label{eq:J0haut}
\mathbf{J}_{0} =
\begin{pmatrix}
-2\sigma_{\Re}  & 0 \\
-2\beta_{\Im}\dfrac{\sigma_{\Re}}{\beta_{\Re}} & 0 \end{pmatrix}\text{ and }
\mathbf{J}_{h} =
\begin{pmatrix}
(2a - 1)|W_{LC}|^{2(a - 1)}\mu_{\Re} & 
-|W_{LC}|^{2(a - 1)}\mu_{\Im} \\
 (2a - 1)|W_{LC}|^{2(a - 1)}\mu_{\Im} & 
|W_{LC}|^{2(a - 1)}\mu_{\Re}
\end{pmatrix}\, .
\end{equation}

Herein, we will consider the case where the underlying structure is a directed graph whose Laplace matrix admits complex eigenvalues, $\Lambda^{(\alpha)}=\Lambda^{(\alpha)}_{\Re}+i\Lambda^{(\alpha)}_{\Im}$. The case of real eigenvalues, i.e., in the case of symmetric networks and some non-normal networks such as directed trees, will thus be obtained by imposing $\Lambda^{(\alpha)}_{\Im}=0$ in the following analysis.  To simplify the notation, let us introduce
\begin{equation}
\label{eq:newvar}
\begin{aligned}
\varphi &= 2a|W_{\mathrm{LC}}|^{2(a-1)}\mu_{\Re}\, , \quad
\gamma = -\left(2\beta_{\Im}\frac{\sigma_{\Re}}{\beta_{\Re}}\mu_{\Im} 
    + 2\sigma_{\Re}\mu_{\Re}\right)|W_{\mathrm{LC}}|^{2(a-1)}, \\
&\text{and }\varepsilon = (2a-1)(\mu_{\Re}^{2} + \mu_{\Im}^{2})|W_{\mathrm{LC}}|^{4(a-1)}\, .
\end{aligned}
\end{equation}

The eigenvalues of~\eqref{eq:Jalphaaut} are easily computed and take the form~\cite{asllani2014theory,di2017benjamin}
\begin{equation} \label{eq:eigenvalues}
    \lambda_{\alpha} = \frac{1}{2} \left[ \left(\mathrm{tr} \mathbf{J}_{\alpha} \right)_{\Re} + \varpi \right] + \frac{1}{2} \left[ \left(\mathrm{tr} \mathbf{J}_{\alpha} \right)_{\Im} + \delta \right]i,
\end{equation}
where $\varpi = \sqrt{\tfrac{A + \sqrt{A^2 + B^2}}{2}}$, $\delta = \operatorname{sign}(B)\sqrt{\tfrac{-A + \sqrt{A^2 + B^2}}{2}}$
and
\begin{equation*}
    A = \left[\left(\mathrm{tr} \mathbf{J}_{\alpha}\right)_{\Re}\right]^2 - \left[\left(\mathrm{tr} \mathbf{J}_{\alpha}\right)_{\Im}\right]^2 - 4\left(\det \mathbf{J}_{\alpha}\right)_{\Re} \text{ and }
    B = 2\left(\mathrm{tr} \mathbf{J}_{\alpha}\right)_{\Re}\left(\mathrm{tr} \mathbf{J}_{\alpha}\right)_{\Im} - 4\left(\det \mathbf{J}_{\alpha}\right)_{\Im}\, .
\end{equation*}
Moreover,
\begin{align*}
  \left(\mathrm{tr}\mathbf{J}_{\alpha}\right)_{\Re}
    &= -2\sigma_{\Re} + \varphi \Lambda_{\Re}^{(\alpha)}
  &\qquad
  \left(\det\mathbf{J}_{\alpha}\right)_{\Re}
    &= \gamma \Lambda_{\Re}^{(\alpha)}
     + \varepsilon \bigl[(\Lambda_{\Re}^{(\alpha)})^2 - (\Lambda_{\Im}^{(\alpha)})^2\bigr], \\
  \left(\mathrm{tr}\mathbf{J}_{\alpha}\right)_{\Im}
    &= \varphi \Lambda_{\Im}^{(\alpha)}
  &\qquad
  \left(\det\mathbf{J}_{\alpha}\right)_{\Im}
    &= \gamma \Lambda_{\Im}^{(\alpha)}
     + 2\varepsilon \,\Lambda_{\Re}^{(\alpha)}\,\Lambda_{\Im}^{(\alpha)}\, .
\end{align*}
We can now define the dispersion relation, namely, the root~\eqref{eq:eigenvalues} with the largest real part, as
\begin{equation}
\label{eq:reldisp}
\lambda (\Lambda^{(\alpha)}) = \max \Re\lambda_\alpha\, ,
\end{equation}
where we emphasized the dependence of the latter on the eigenvalues of the Laplace matrix. Following closely the analysis proposed in~\cite{asllani2014theory,di2017benjamin}, we can prove the existence of two polynomials\footnote{Let us observe that we hereby follow the notation proposed in~\cite{di2017benjamin}, which slightly differs from the one introduced in~\cite{asllani2014theory}.}
\begin{equation*}
    S_1\left(x\right) = C_{12}x^2 + C_{11}x + C_{10}\text{ and }
    S_2\left(x\right) = C_{24}x^4 + C_{23}x^3 + C_{22}x^2 + C_{21}x \, ,
\end{equation*}
whose coefficients are explicitly given by
\begin{align*}
  C_{24} &= \varphi^2 \varepsilon,
  &\;
  C_{12} &= 4\varepsilon^2 - \varphi^2 \varepsilon, \\
  C_{23} &= \varphi^2 \gamma - 4\varphi\sigma_{\Re} \varepsilon,
  &\;
  C_{11} &= 4\gamma\varepsilon - \varphi^2 \gamma, \\
  C_{22} &= 4\sigma_{\Re}^2 \varepsilon - 4\varphi\sigma_{\Re} \gamma,
  &\;
  C_{10} &= \gamma^2 + 2\varphi\sigma_{\Re}\gamma + 4\sigma_{\Re}^2 \varepsilon, \\
  C_{21} &= 4\sigma_{\Re}^2 \gamma\,.
\end{align*}
Then, the dispersion relation~\eqref{eq:reldisp} is positive for some $\Lambda^{(\alpha)}$ if and only if
\begin{equation}
\label{eq:condS2S1}
S_{2}\left(\Lambda_{\Re}^{(\alpha)}\right)  < \left( \Lambda_{\Im}^{(\alpha)} \right)^2  S_{1}\left(\Lambda_{\Re}^{(\alpha)}\right) \, .
\end{equation}
Stated differently, the condition $S_2(x) < y^2S_1(x)$ defines a region in the complex plane (instability region) and the instability of the synchronous solution amounts to requires the existence of one complex eigenvalue $\Lambda^{(\alpha)}=\Lambda^{(\alpha)}_\Re+ i\Lambda^{(\alpha)}_\Im$ lying in such region.

\begin{remark}[The case of real eigenvalues of the Laplace matrix]
 Let us briefly consider the case of real eigenvalues of the Laplace matrix holding true for symmetric networks and some asymmetric ones as well. In this case the previous formulas simplify and the dispersion relation~\eqref{eq:reldisp} rewrites
\begin{equation} 
\label{GrowthratesGen}
    \lambda(x) = \frac{1}{2}\left[-2\sigma_{\Re} - \varphi x + \sqrt{4\sigma_{\Re}^{2} + 4(\sigma_{\Re}\varphi + \gamma)x + (\varphi^{2} - 4\varepsilon)x^{2}}\right]\, ,
\end{equation}
where we set $x = -\Lambda^{(\alpha)}\geq 0$. Let us observe that we always have $\lambda(0)=0$, while a second root is given by
\begin{equation}
\label{eq:x2}
 x_2=\frac{\gamma}{\varepsilon}= -\frac{2}{2a-1}\frac{\beta_{\Im}\frac{\sigma_{\Re}}{\beta_{\Re}}\mu_{\Im} + \sigma_{\Re}\mu_{\Re}}{(\mu_{\Re}^{2} + \mu_{\Im}^{2})|W_{\mathrm{LC}}|^{2(a-1)}}\, .
\end{equation}
Under the assumption of dealing with positive integers $a$ and $b$, we have $\varepsilon >0$, which implies that the sign of $x_2$ is fully determined by $\gamma$. Looking for negative $x_2$ (remember that $x=-\Lambda^{(\alpha)}$) implies thus assuming $\gamma <0$. Because $\lambda(x)$ admits two roots, one of which is always $x=0$, this setting implies that $\lambda(x)<0$ for all $x>0$ and thus $\lambda(\Lambda^{(\alpha)})<0$ for all $\alpha=2,\dots,N$. The synchronous solution $W_j(t)=W_{\mathrm{LC}}(t)$, $j=1,\dots,N$, is then stable and the system admits complete synchronization for any network realizing the coupling among the oscillators (see left column of Fig.~\ref{fig:AutoUnidrectednetwork}, where the used parameters return $\gamma=-3$). On the other hand, if $\gamma >0$, we obtain $x_2>0$ and the curve $\lambda(x)$ grows from $0$ until it reaches a maximum, to then decreasing and becoming negative for large $x$. This implies that there exist networks whose spectrum of the Laplace matrix returns a positive dispersion relation preventing the system from achieving synchronization~\footnote{Let us observe that a similar, but weaker, result can be obtained by looking at the behavior of $\lambda(x)$ close to $x=0$. In fact, by computing its derivative, we get $\lambda(x) \approx \frac{\gamma}{2\sigma_{\Re}}x + \mathcal{O}(x^2)$ and thus if $\gamma>0$, the function is locally increasing from zero to positive values.}.

Let us observe that $\gamma$ depends on the exponent $a$ via the term $|W_{\mathrm{LC}}|^{2(a-1)}$, hence the sign of $\gamma$ is the same regardless of the value of $a$. We have thus found an interesting result: in the case of symmetric network, or asymmetric one but with real eigenvalues of the Laplace matrix~\cite{entropy}, the existence of complete synchronization does not depend on the values of $a$ and $b$ (recall that we have imposed $a=b+1$). Stated differently, if the curve $\lambda(x)$ is negative for all $x>0$, the system completely synchronizes with a linear coupling, then it does for any nonlinear one, as long as $a=b+1$. Let us observe that the contrary may not hold true because of finite-size effects. This phenomenon is reported in the middle and right columns of Fig.~\ref{fig:AutoUnidrectednetwork} in the case of an Erd\H{o}s-R'enyi symmetric network composed of $N=150$ nodes and with probability $p=0.03$ to create a link between two distinct nodes~\cite{erdosreny}. In the former case $a=1$, $\gamma=1$ and $x_2=0.5$, while in the latter case $a=5$, $\gamma=1/32$ and $x_2\sim 0.88$. Because the smaller non-zero eigenvalue is $-\Lambda^{(2)}\sim 0.5911$, we clearly have $-\Lambda^{(2)} > 0.5$ and thus $\lambda(\Lambda^{(\alpha)})<0$ for all $\Lambda^{(\alpha)}$, $\alpha =2,\dots, N$. On the other hand, $-\Lambda^{(2)} < 0.88$ and there exists several eigenvalues for which $\lambda(\Lambda^{(\alpha)})>0$. 
\begin{figure}[h!]
    \centering
    \includegraphics[width=0.3\linewidth]{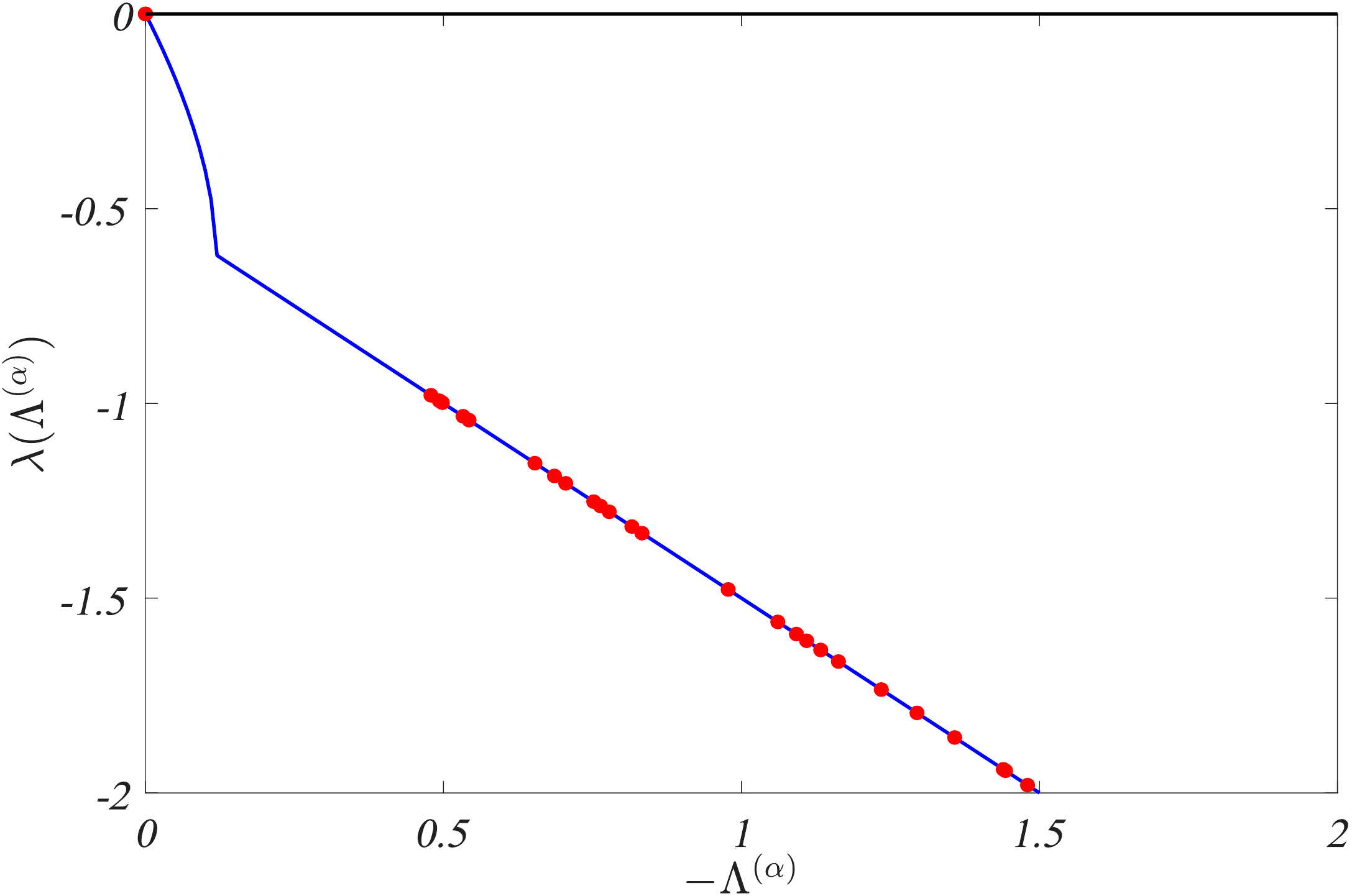}  \hfill \includegraphics[width=0.3\linewidth]{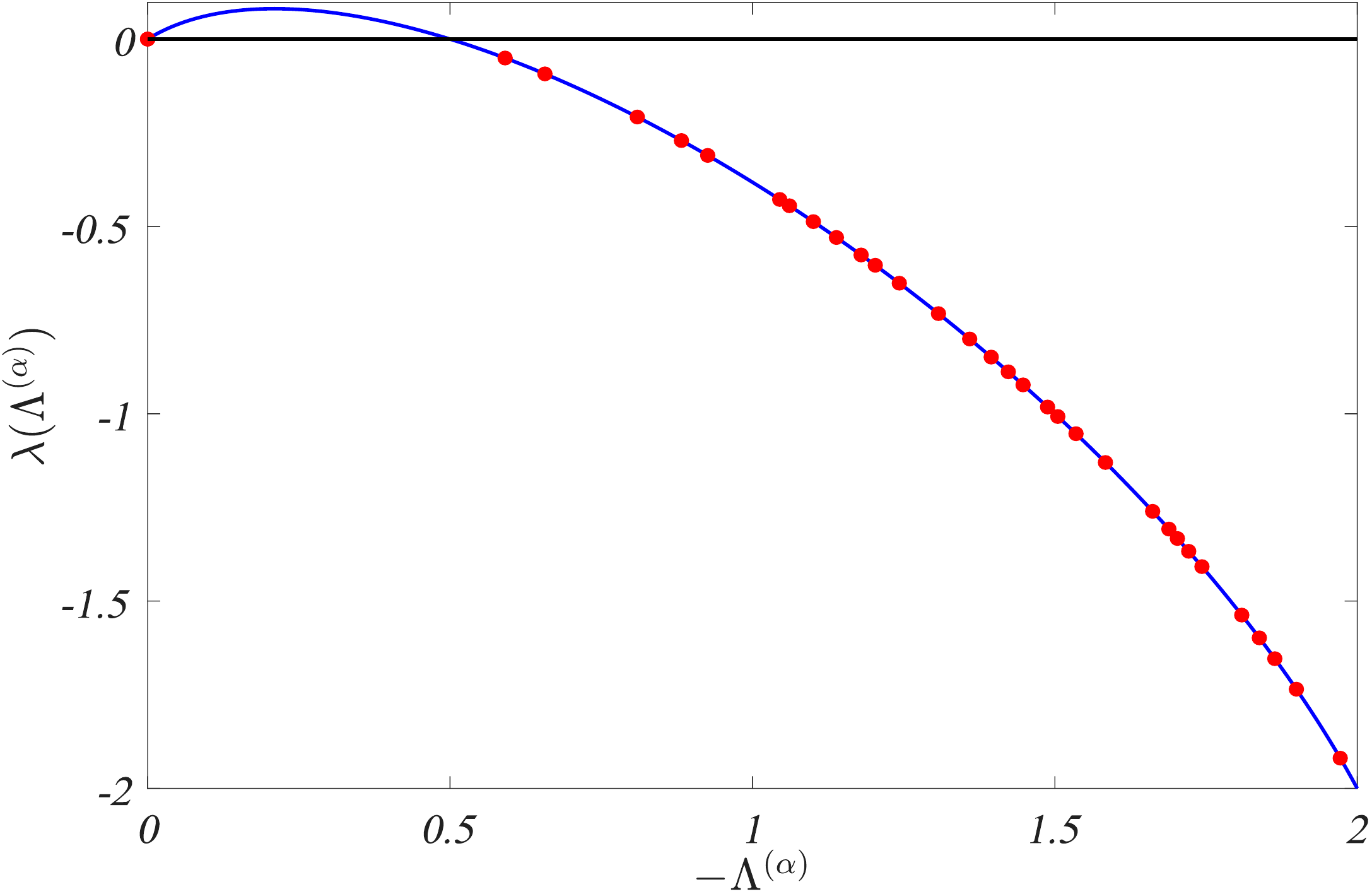}   \hfill \includegraphics[width=0.3\linewidth]{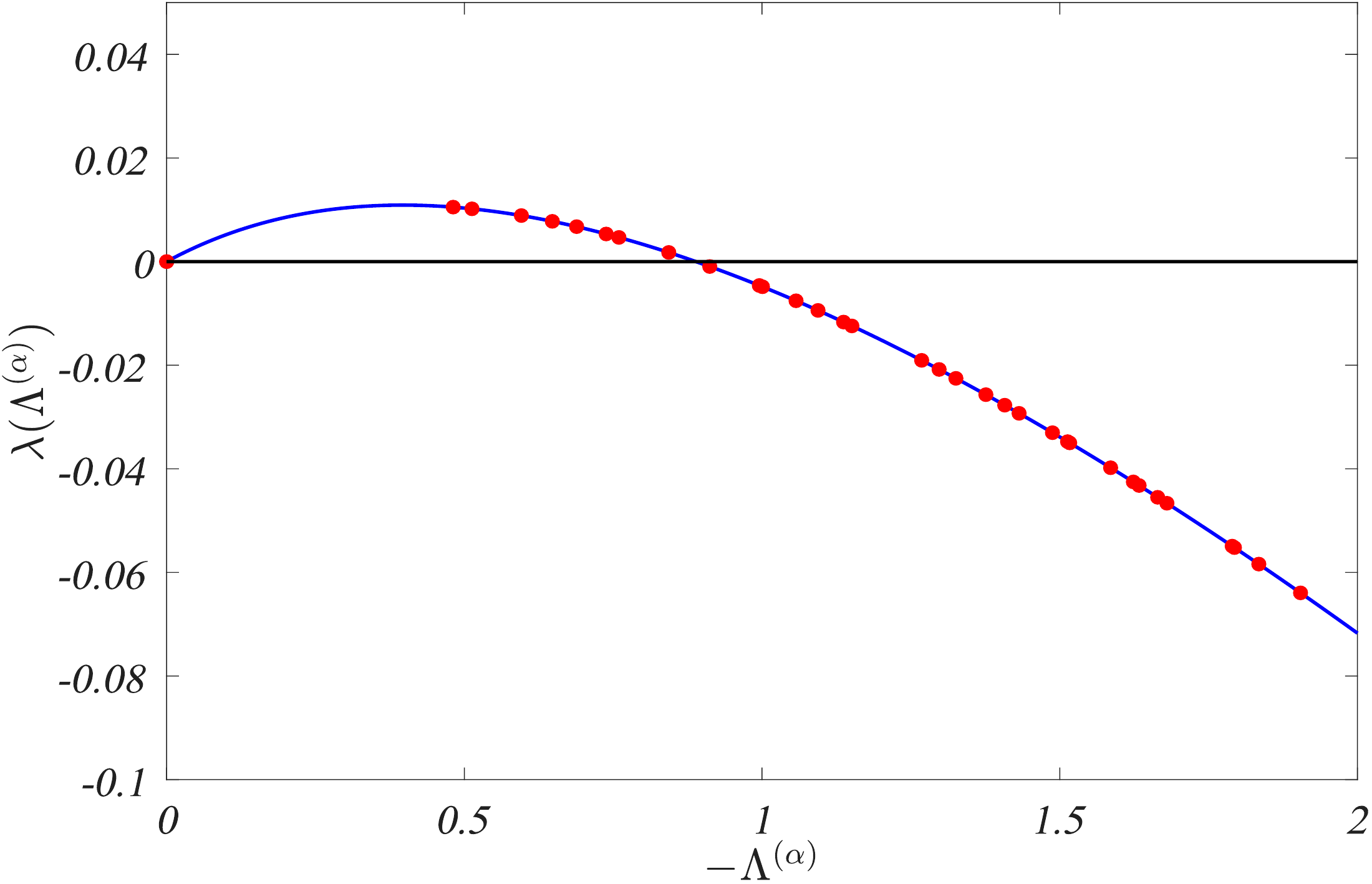}   \vspace{1em}\\
    \includegraphics[width=0.3 \linewidth, height=2.8cm]{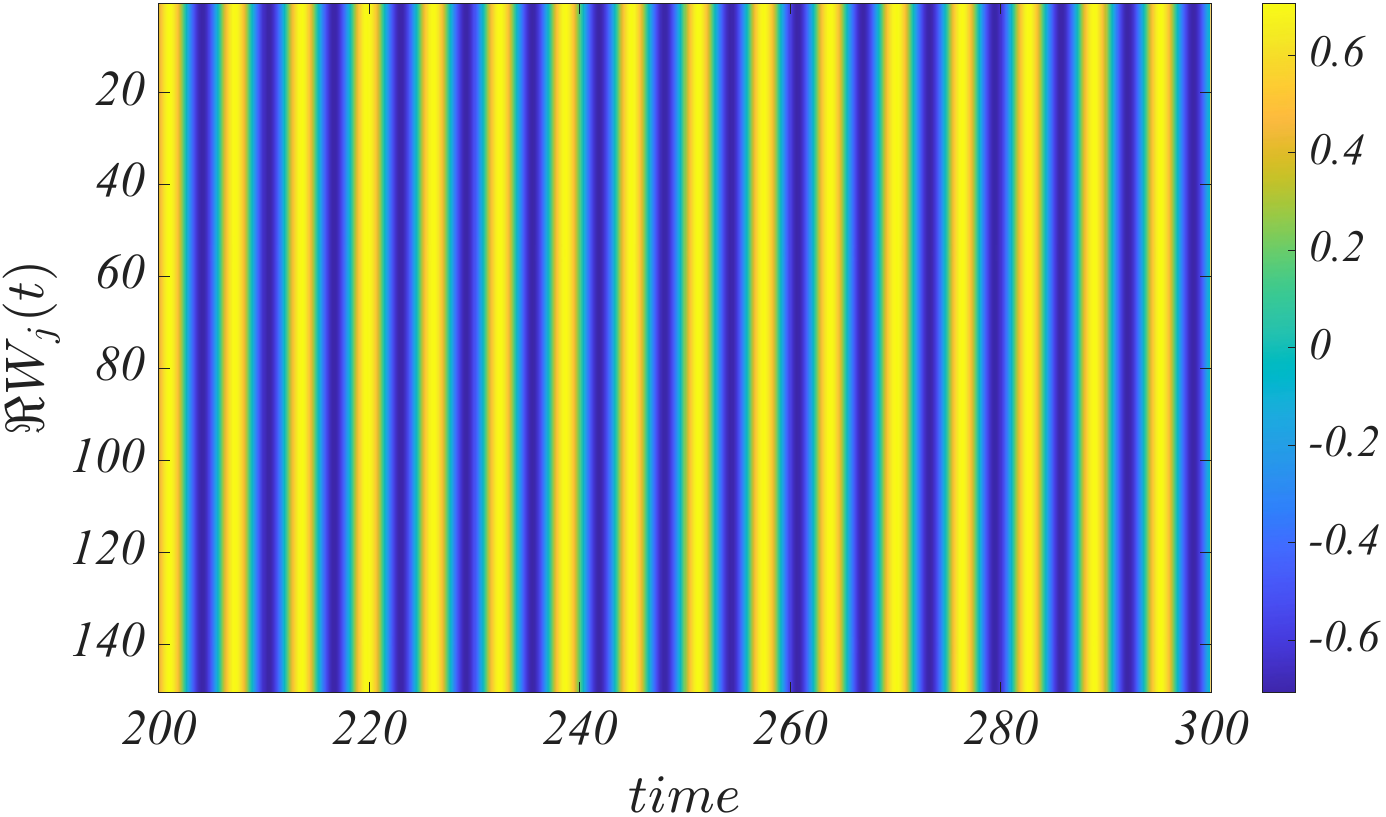}\hfill     \includegraphics[width=0.3 \linewidth, height=2.8cm]{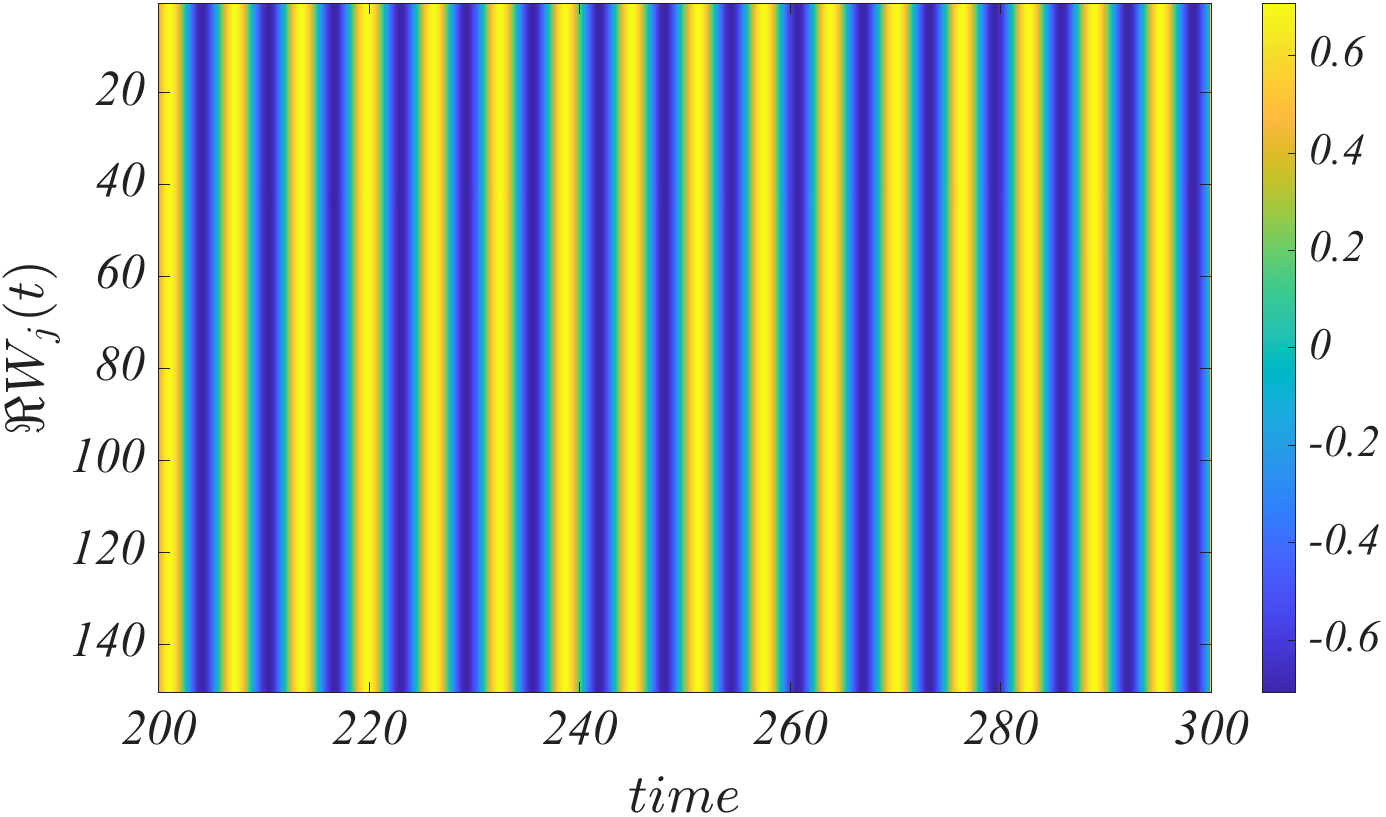}\hfill \includegraphics[width=0.3 \linewidth, height=2.8cm]{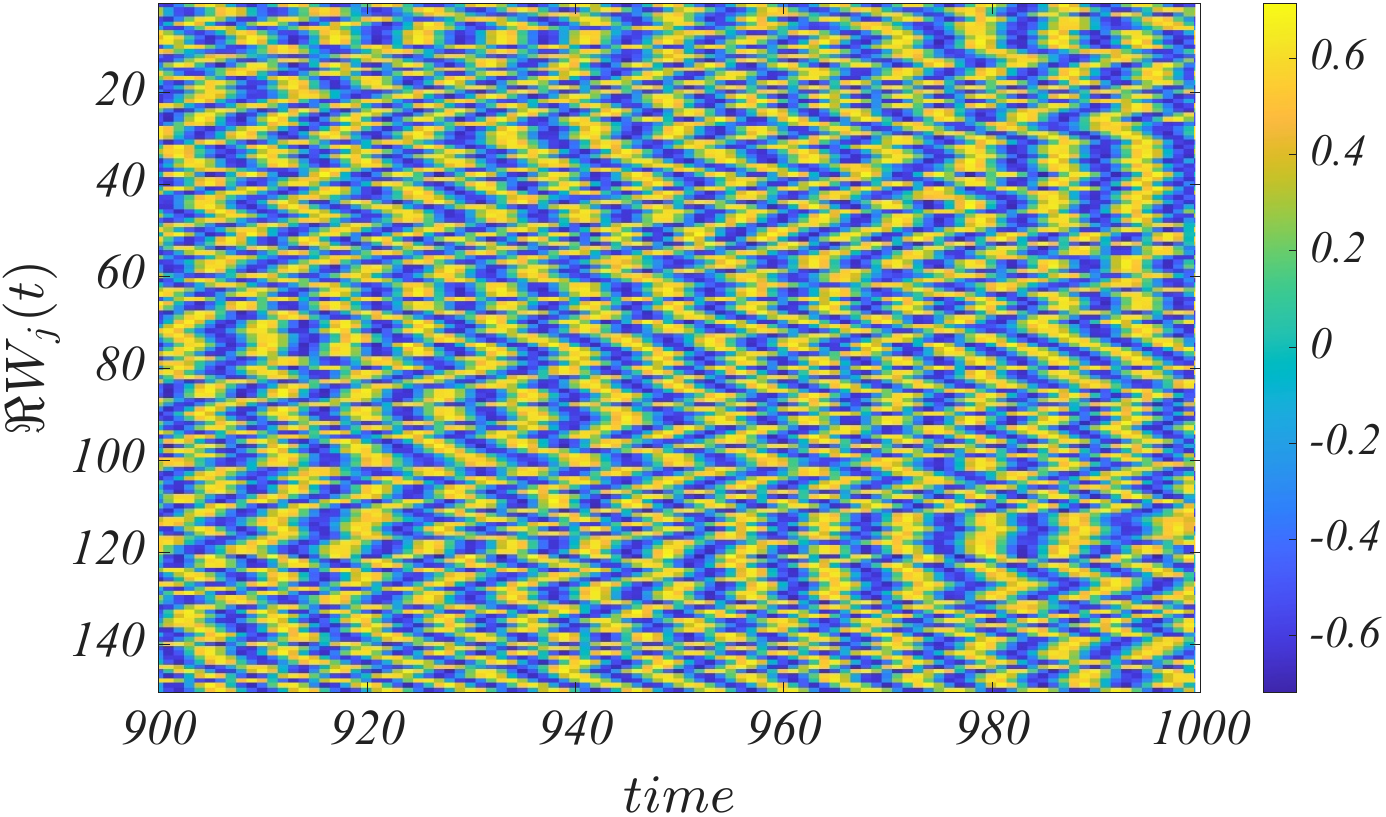}
    \caption{
    \textbf{Complete synchronization of Stuart-Landau oscillators coupled with a symmetric network, $a=b+1$.} The left column corresponds to $\mu=1+i$, $a=1$ and $b=0$, the dispersion relation (blue curve and red dots, top panel) is negative and the oscillators synchronize as we can appreciate by looking at the time evolution of $\Re W_j(t)$ (bottom panel). The middle column shows the results for to $\mu=1-i$, $a=1$ and $b=0$, the curve $\lambda(x)$ is positive (blue curve, top panel) however the  dispersion relation is negative (red dots, top panel) and the system still exhibits synchronization as testified by the behavior of $\Re W_j(t)$ (bottom panel). The right column represents the case $\mu=1-i$, $a=5$ and $b=4$, both the curve $\lambda(x)$ and the dispersion relation are positive (blue curve and red dots, top panel) and the system is not able to exhibit synchronization as shown by the time evolution of $\Re W_j(t)$ (bottom panel). The coupling is represented by an Erd\H{o}s–R\'enyi symmetric network composed of $N=150$ nodes and a probability $p=0.03$ for an edge to exist among any couple of nodes. The remaining model parameters are $\sigma=0.5$ and $\beta=1+2i$.}
    \label{fig:AutoUnidrectednetwork}
\end{figure}


The reason for this behavior is due to the fact that $x_2$ depends on $a$ (see Eq.~\eqref{eq:x2}): if $|W_{\mathrm{LC}}|\geq 1$, then $x_2$ decreases with increasing values of $a$, and thus the interval for which $\lambda(x)$ is positive shrinks if $a$ increases. This reduces the probability that the spectrum of the Laplace matrix of a ``generic'' network falls inside this interval, increasing the probability for the system to synchronize completely. On the other hand, if $|W_{\mathrm{LC}}|< 1$, then $x_2$ decreases for $a \in [0,\hat{a}]$, with $\hat{a}=\left(1-1/\log (|W_{\mathrm{LC}}|)\right)/2$, where it reaches its minimum and then increases unbounded. In conclusion by increasing $a$, the size of the interval for which $\lambda(x)>0$ first decreases and then increases, hence the probability that the eigenvalues of the Laplace matrix of a generic network lie in this region is thus large if $a$ is large. Stated equivalently, the probability of the system to not achieve synchronization increases with large $a$.
\begin{figure}[h!]
    \centering
     \includegraphics[width=0.4\linewidth]{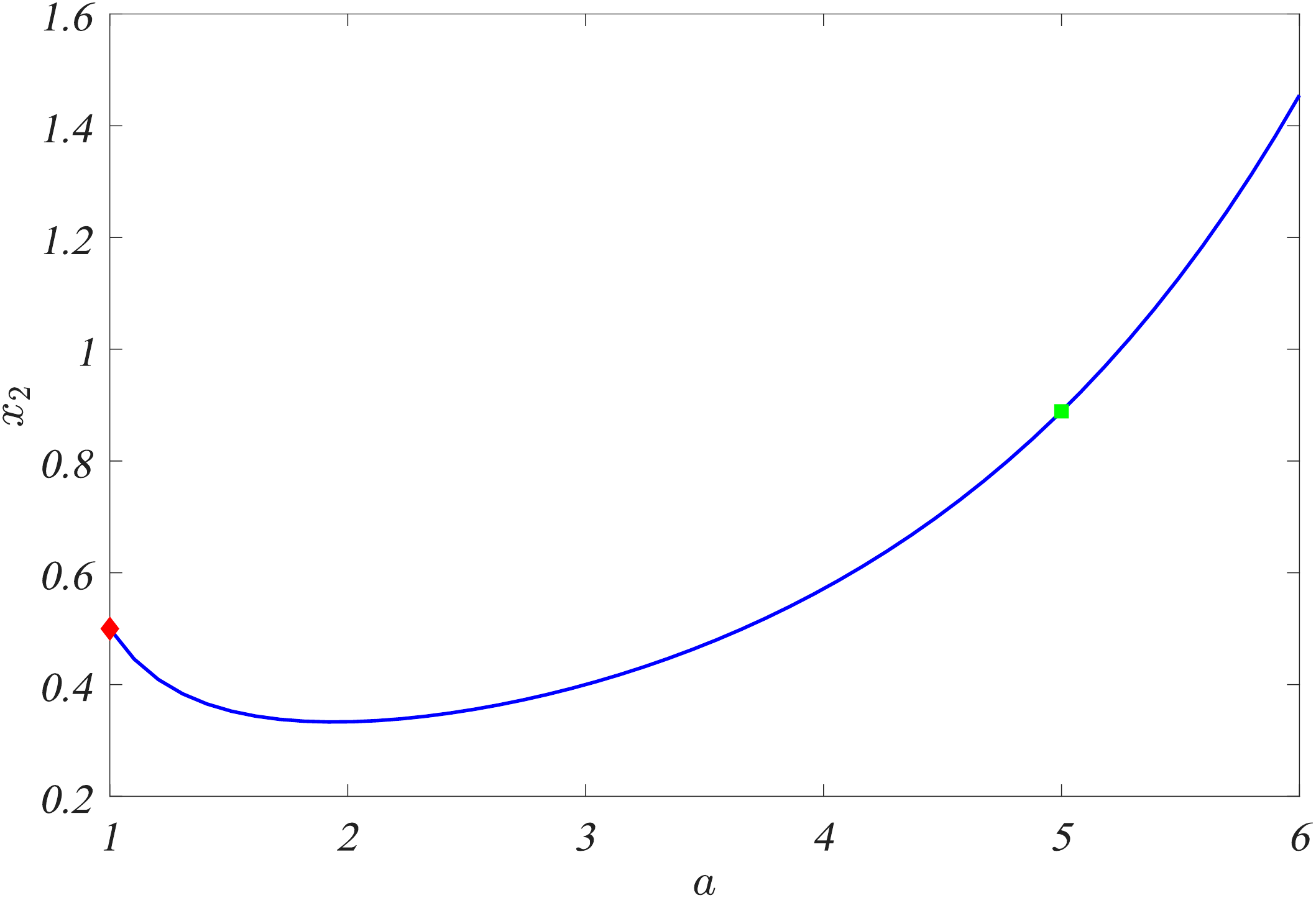}
    \caption{
    \textbf{The dependence on the parameter $a$ of the nonzero root, $x_2$, of $\lambda(x)$.} By using the parameters selected in the middle and right column of Fig.~\ref{fig:AutoUnidrectednetwork}, i.e., $\sigma=0.5$, $\beta=1+2i$ and $\mu=1-i$, we show the variation of $x_2$ as a function of $a$. Because $|W_{\mathrm{LC}}|=1/\sqrt{2}<1$ the function exhibits a non-monotone behavior as mentioned in the text. We emphasized two values, $x_2=1/2$ for $a=1$ (red diamond) and $x_2\sim 0.88$ for $a=5$ (green square).}
    \label{fig:x2}
\end{figure}
\end{remark}

One of the main consequences of Eq.~\eqref{eq:condS2S1} is that, as expected from previous works~\cite{asllani2014theory,di2017benjamin,entropy},  directed networks can prevent synchronization to emerge, because of the presence of complex eigenvalues of the Laplace matrix. By assuming linear coupling, i.e., $a=1$ and $b=0$, it has been shown~\cite{di2017benjamin} that, if the condition $\beta_\Im \mu_\Im+\beta_\Re\mu_\Re>0$ is satisfied, then instabilities may\footnote{As discussed earlier, the instability conditions are necessary but not sufficient due to finite-size effects: in fact, the conditions could be satisfied, but in the interval of positive dispersion relation there are no eigenvalues of the Laplace matrix.} arise due to the contribution of the imaginary part of the eigenvalues, while in the case of real eigenvalues the system would completely synchronize. Let us note that this condition corresponds to $\gamma <0$, which is often found in the literature, see, e.g.,~\cite{di2017benjamin}, as $1+c_1c_2>0$ when authors use the notation $\beta=1+ic_2$ and $\mu=1+ic_1$. Here, we demonstrated that this claim remains valid even though the coupling function is nonlinear, as long as the condition $a=b+1$ holds true.

\begin{figure}[h!]
    \centering 
    \includegraphics[width=0.3\linewidth]{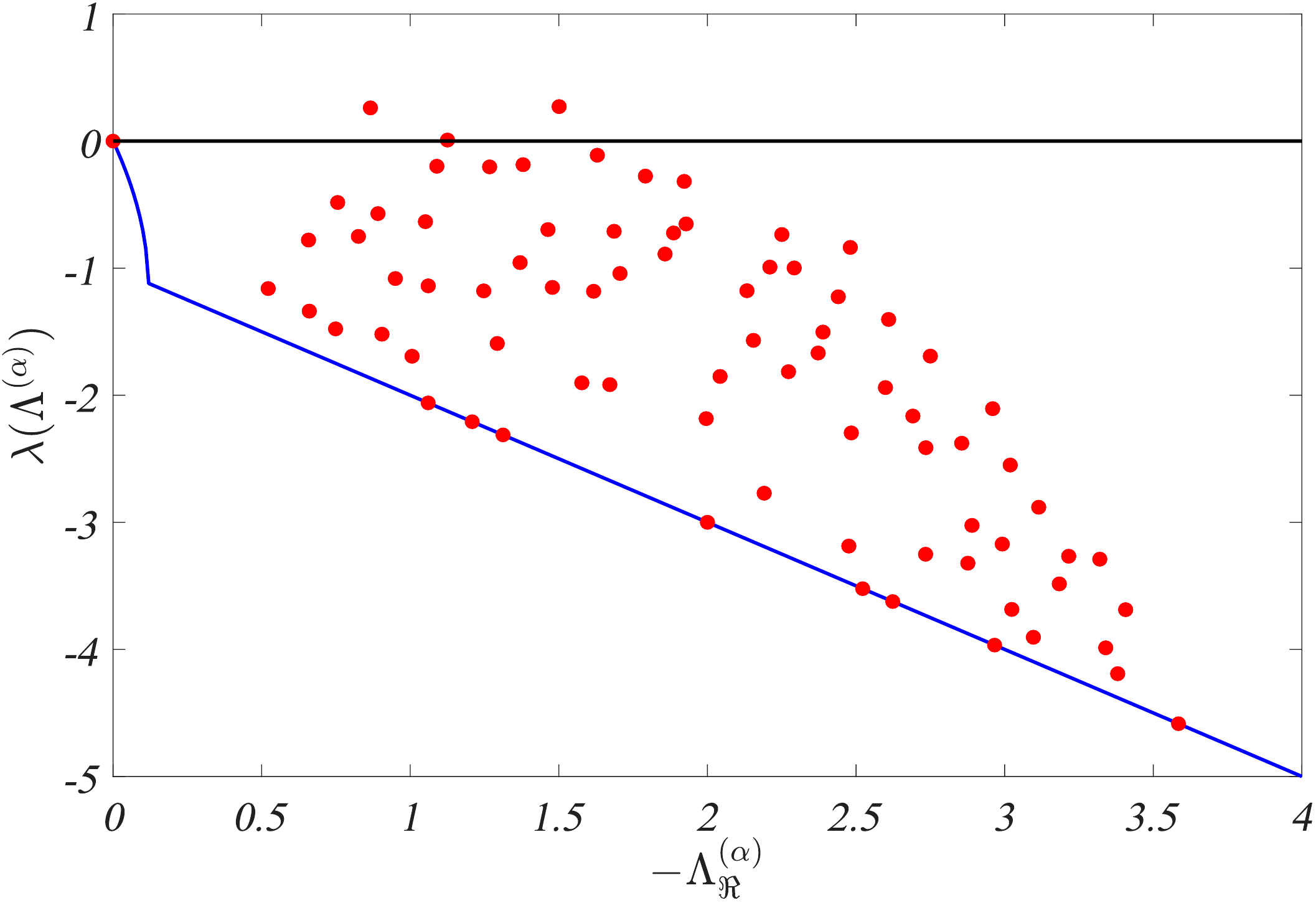} \quad   \includegraphics[width=0.3\linewidth]{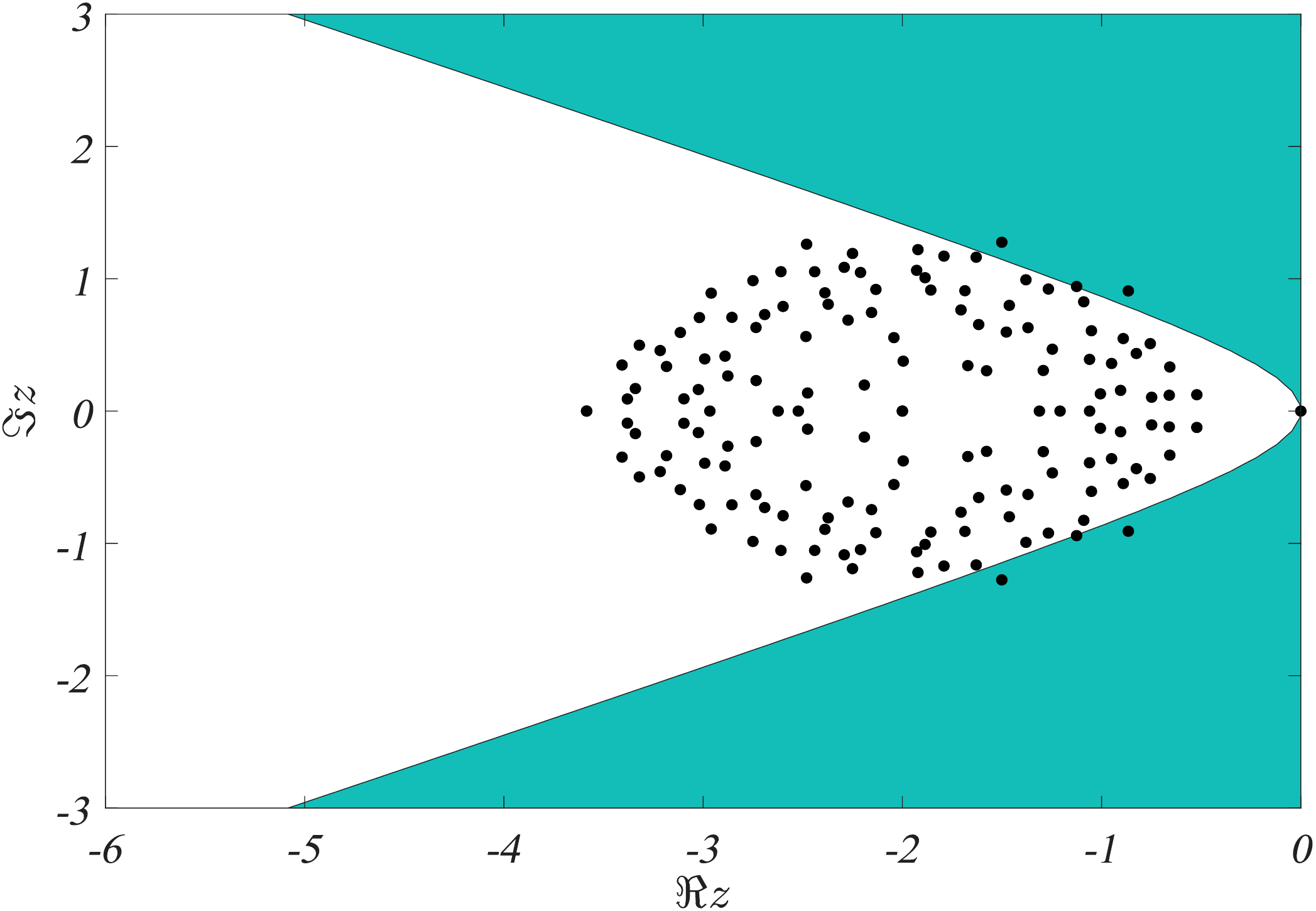}\quad     \includegraphics[width=0.3 \linewidth, height=2.8cm]{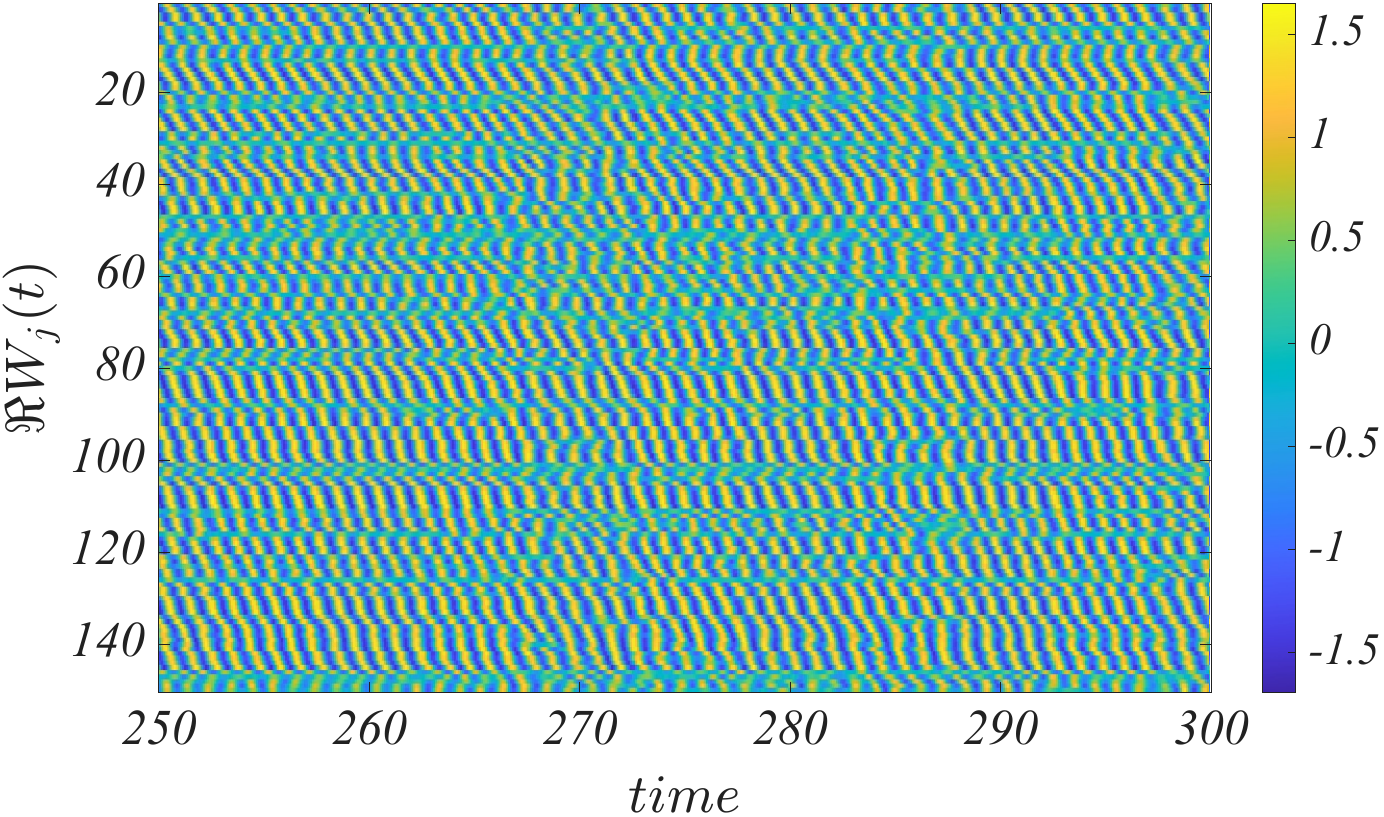}\vspace{1em}\\
        \includegraphics[width=0.3\linewidth]{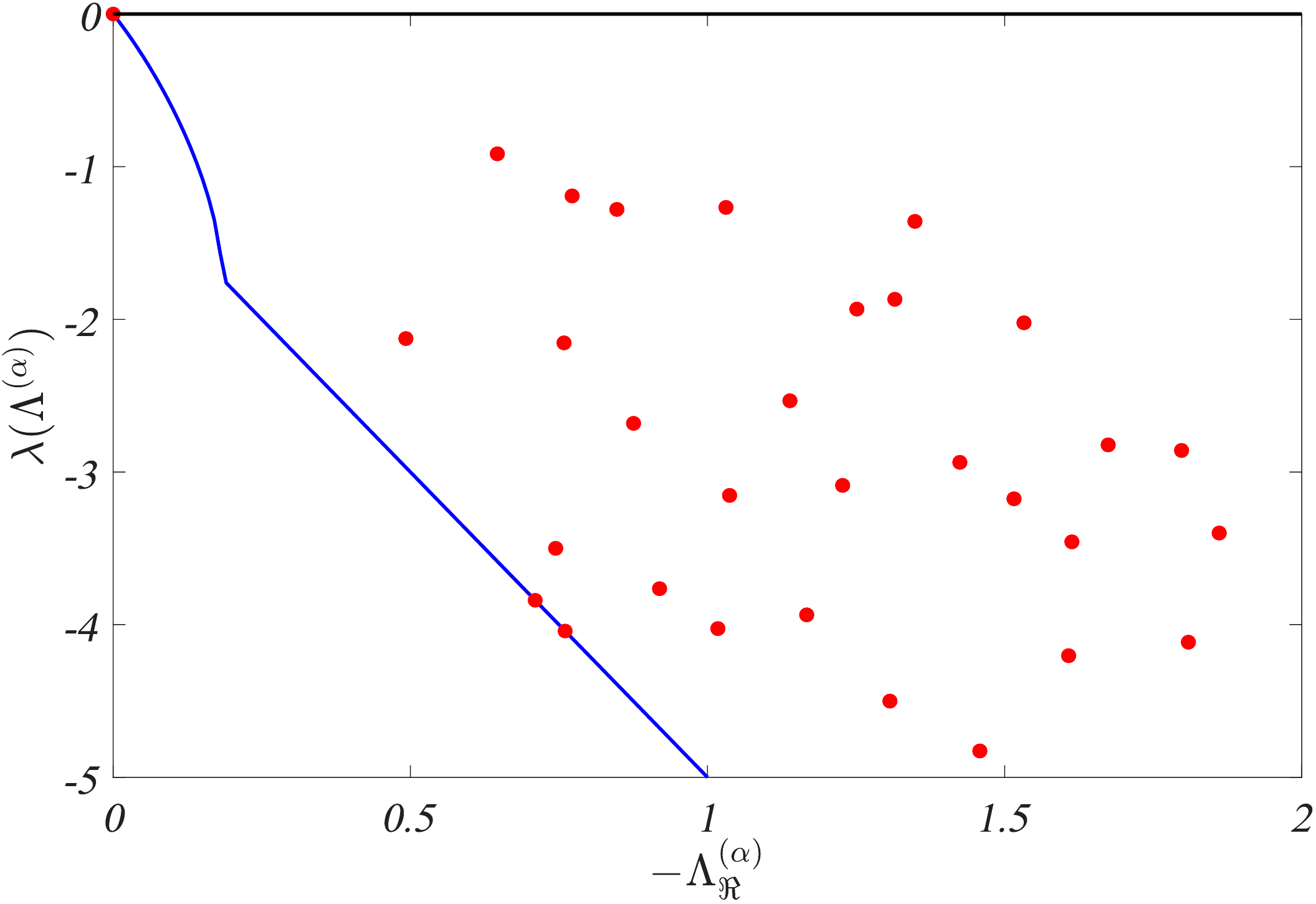}\quad        \includegraphics[width=0.3 \linewidth]{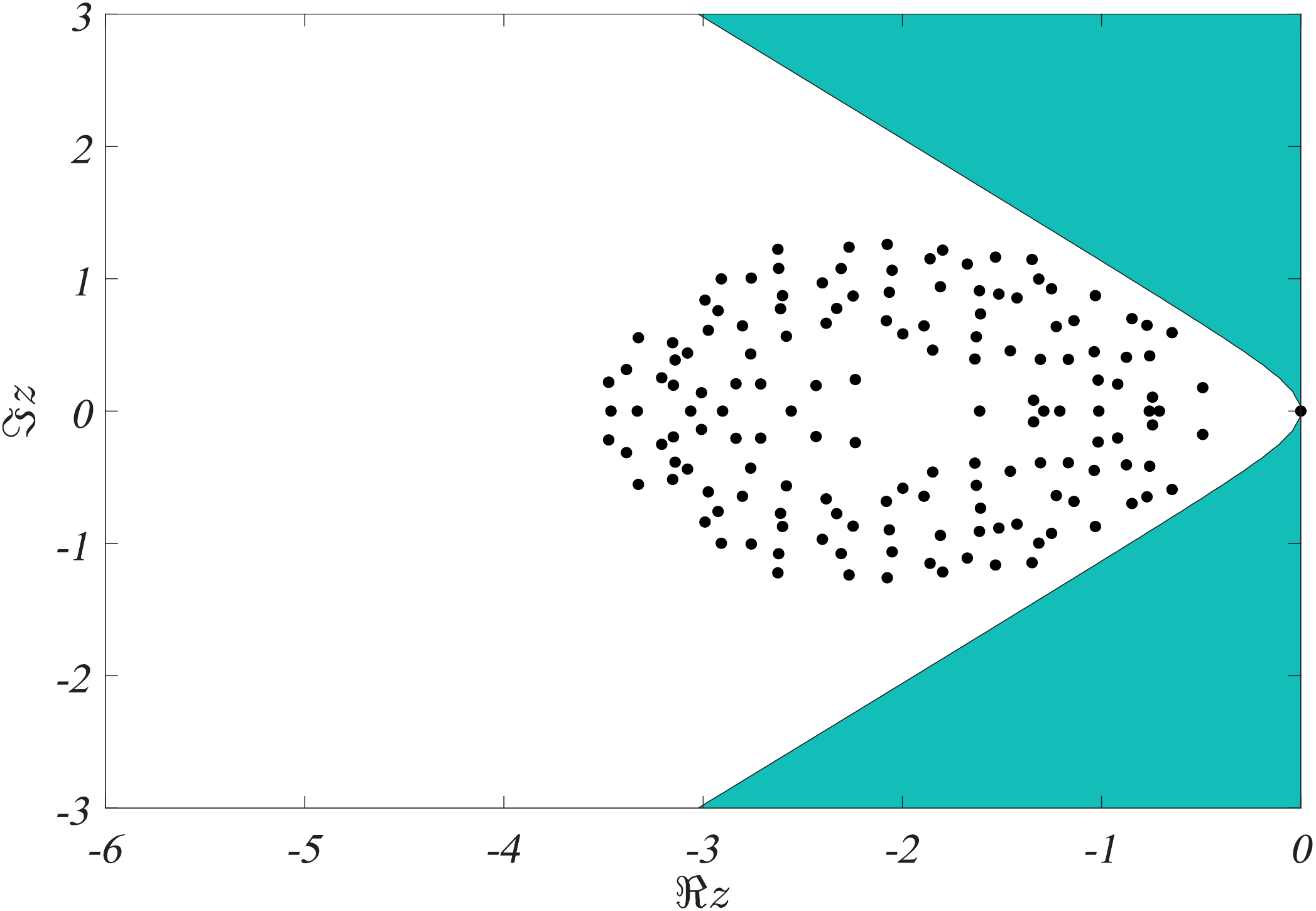}\quad     \includegraphics[width=0.3 \linewidth, height=2.8cm ]{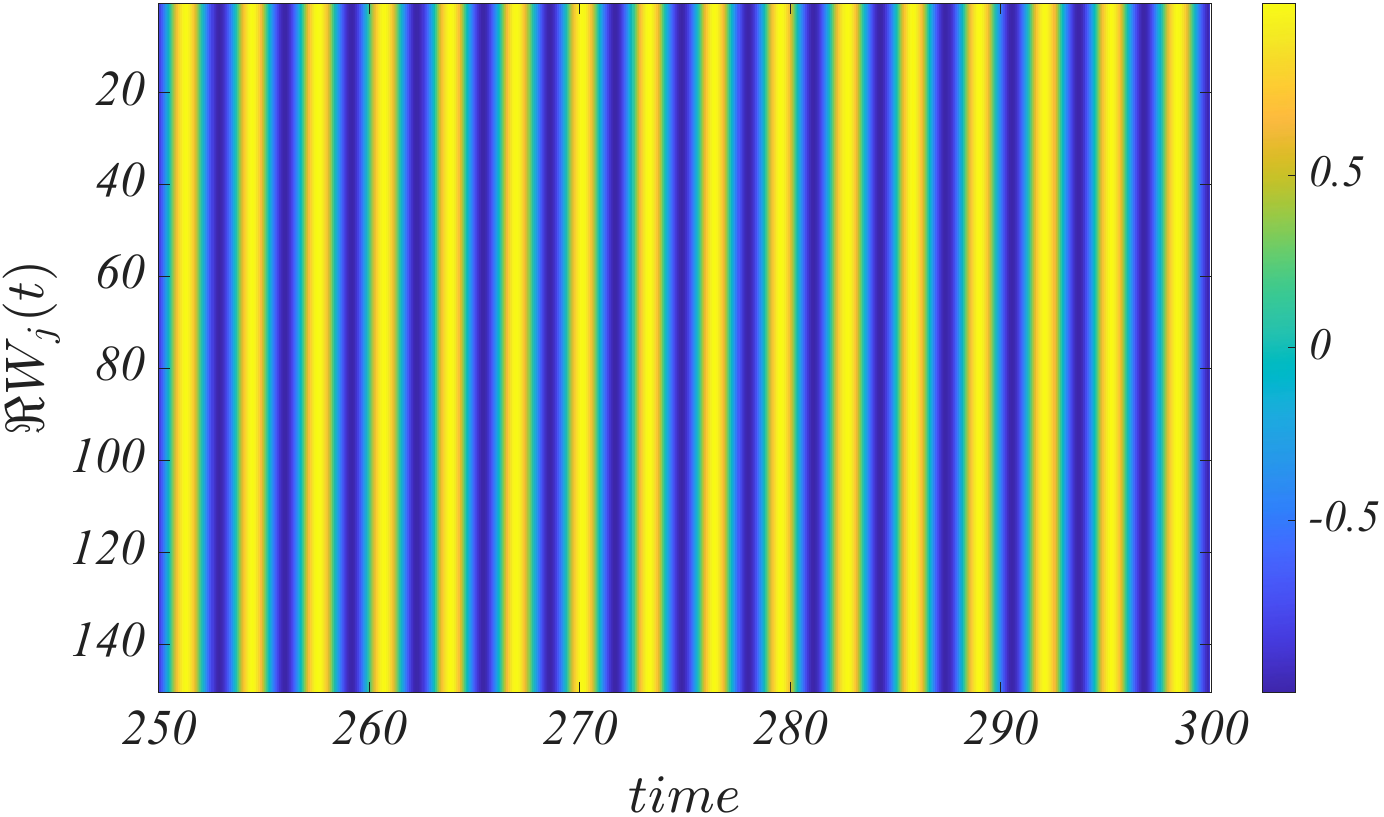}
    \caption{
    \textbf{Complete synchronization of Stuart-Landau oscillators coupled via a directed network, $a=b+1$.} We show the results of SL oscillators coupled via a directed network composed of $N=150$ nodes obtained by using the Watts-Strogatz algorithm with $p_{WS}=0.9$ the probability to rewire any directed network starting from a undirected ring where each node has degree $2$. The model parameters have been set to $\sigma = 1.0$, $\beta=1+2i$ and $\mu=1+2i$; moreover, $a=1$ (top row) and $a=4$ (bottom row). Panels in the left column report the curve $\lambda(x)$ (blue) and the dispersion relation $\lambda(\Lambda^{(\alpha)})$ (red dots). It can be observed that now the latter deviates from the curve, this being a signature of the presence of a nonzero imaginary part of the spectrum of the Laplace matrix. Moreover, in the top panel, the dispersion relation assumes positive values, while this does not happen for the parameters corresponding to the bottom panel. The middle columns represent the region of the complex plane where the condition $S_2(x) < y^2S_1(x)$ is satisfied (green region). It can be observed that, in the top left panel, some complex eigenvalues (black dots) belong to the instability region, thus impeding the system to synchronize, as shown in the bottom left panel, where we report $\Re W_j(t)$. By increasing $a$, we can observe that the instability region shrinks (top right panel) and all the complex eigenvalues (black dots) fall outside the instability region; this allows complete synchronization (see bottom right panel).}
   \label{fig:Autodrectednetwork}
\end{figure}

To support this conclusion, we considered a directed network obtained starting from a $2$-regular ring and applying the Watts-Strogatz algorithm, i.e., with probability $p_{WS}=0.9$, each directed link is rewired avoiding multiple links and self-loops~\cite{watts1998collective}. The results are reported in Fig.~\ref{fig:Autodrectednetwork}. The panels in the left columns depict the curve $\lambda(x)$ (blue curve) and the  dispersion relation $\lambda(\Lambda^{(\alpha)})$ (red dots). The latter does not lie on the curve because of the presence of complex eigenvalues of the Laplace matrix. This can allow a positive dispersion relation even if the curve $\lambda(x)$ is negative, which means that the system does not synchronize because of the directionality. A complementary view can be obtained from the panels in the middle column, where we show the instability region in the complex plane defined by the inequality~\eqref{eq:condS2S1} (green regions). For $a=1$ (top row), the instability region is large enough to contain some of the complex eigenvalues (black dots), meaning that we cannot obtain complete synchronization as confirmed by the time evolution of $\Re W_j(t)$ (top right panel). Let us observe that the used parameter values satisfy the condition $\beta_\Im \mu_\Im+\beta_\Re\mu_\Re>0$. In the case of real eigenvalues, we have been able to explicitly show the impact of the nonlinearity via the parameter $a$. A similar conclusion in the case of complex spectrum is difficult to obtain. We hereby propose an example where a stronger nonlinearity, $a=5$, shrinks the instability region so that all complex eigenvalues (black dots) lie outside of it (see bottom left and middle panels of Fig.~\ref{fig:Autodrectednetwork}). In this case, the system achieves complete synchronization (right bottom panel).

\subsection{The non-resonant case: Non-autonomous linear equation}
\label{ssec:nonauton}

Let us now consider the general case $a\neq b+1$, but still $a$ and $b$ to be positive integers. It then follows that the matrix $\mathbf{J}_{\alpha}(t)$ defined in~\eqref{eq:SLLinearmodelGraphTimDep} is time dependent and more precisely it is $T$-periodic, $T=2\pi/[(a-b-1)\omega]$. The stability of the solution $W_j(t)=W_{\mathrm{LC}}(t)$ for all $j=1,\dots,N$ can be proved by resorting to Floquet theory~\cite{Verhulst1996,strogatz2018nonlinear}. This amounts to solve the matrix ordinary differential equation
\begin{equation}
\label{eq:Mode}
 \frac{d\mathbf{M}_{\alpha}}{dt}=\mathbf{J}_{\alpha}(t)\mathbf{M}_{\alpha}\, ,
\end{equation}
where the $2\times 2$ unknown fundamental matrix $\mathbf{M}_{\alpha}$ satisfies the initial condition $\mathbf{M}_{\alpha}(0)=\mathbf{I}_2$, i.e., the identity matrix. From the matrix $\mathbf{M}_{\alpha}(t)$ we can compute the monodromy matrix $\mathbf{C}_{\alpha} := \mathbf{M}_{\alpha}(T)$, whose eigenvalues $\nu_{(\alpha)}$, called the \textit{Floquet multipliers} allow to state about the stability of the synchronous solution. More precisely, after defining the \textit{Floquet exponents}, $\zeta_{\alpha} = \frac{1}{T} \log \nu_{(\alpha)}$, we can prove that the solution $W_j(t)=W_{\mathrm{LC}}(t)$ for all $j=1,\dots,N$ is asymptotically stable if all Floquet exponents, but one that has zero real part, have strictly negative real parts. We can thus define, analogously to the dispersion relation~\eqref{eq:reldisp}, the maximum Floquet exponent as
\begin{equation}
\label{eq:maxFloq}
\zeta (\Lambda^{(\alpha)}) = \max \Re\zeta_\alpha\, .
\end{equation}
Let us observe that the latter is a particular case of the Master Stability Function, whose validity goes beyond the present case of periodic non-autonomous system, allowing to deal with, e.g., synchronization of chaotic systems. Similarly to what we have done in Section~\ref{ssec:autonom}, with a slight abuse of notation, we define $\zeta (x)$ as the maximum Floquet exponent once we replace $\Lambda^{(\alpha)}$ with the continuous variable $x=-\Lambda^{(\alpha)}$.

\begin{figure}[h!]
    \centering 
       \includegraphics[width=0.4\linewidth]{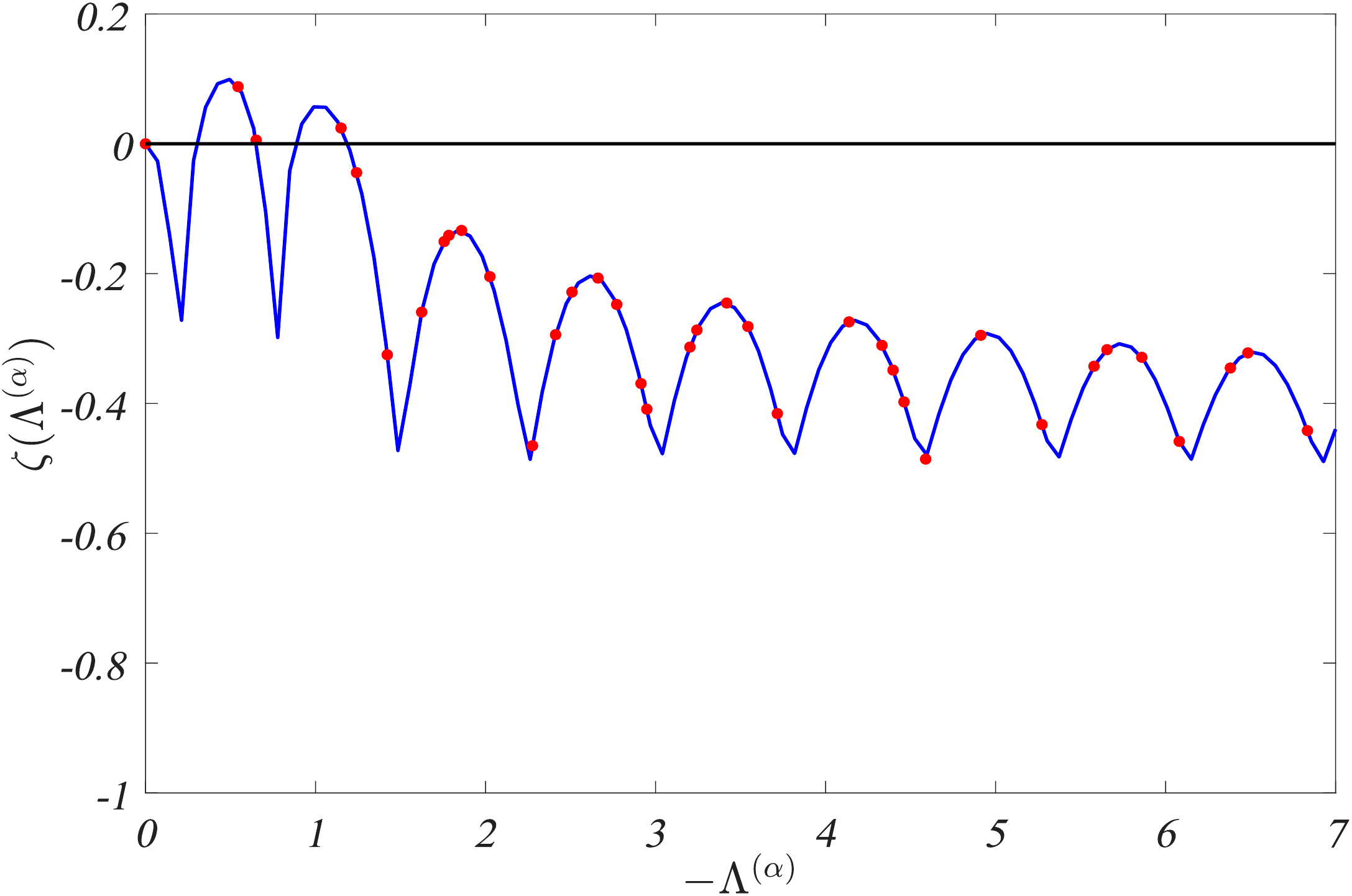}\quad     \includegraphics[width=0.4\linewidth]{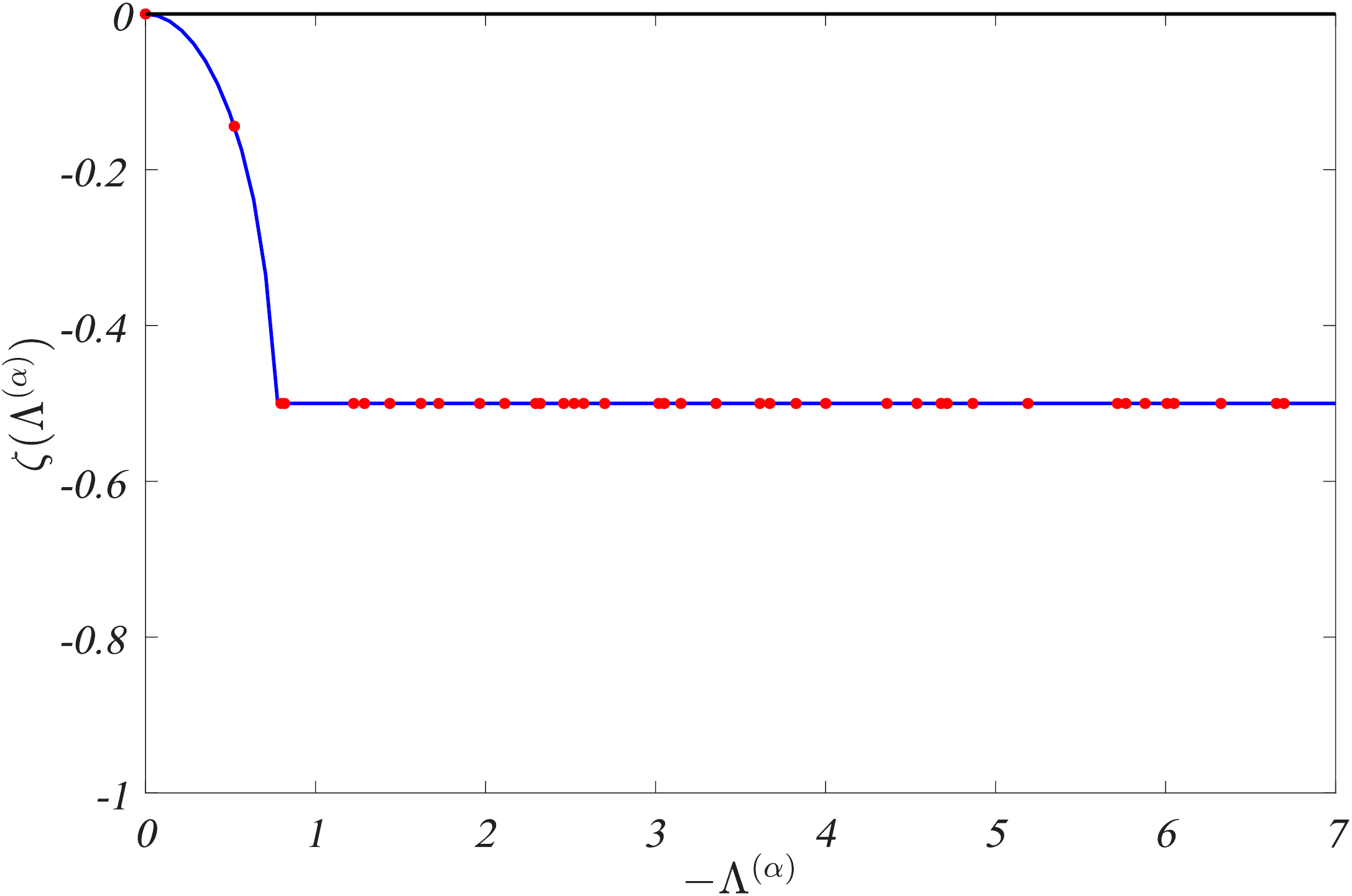}\vspace{1em}\\
        \includegraphics[width=0.4\linewidth]{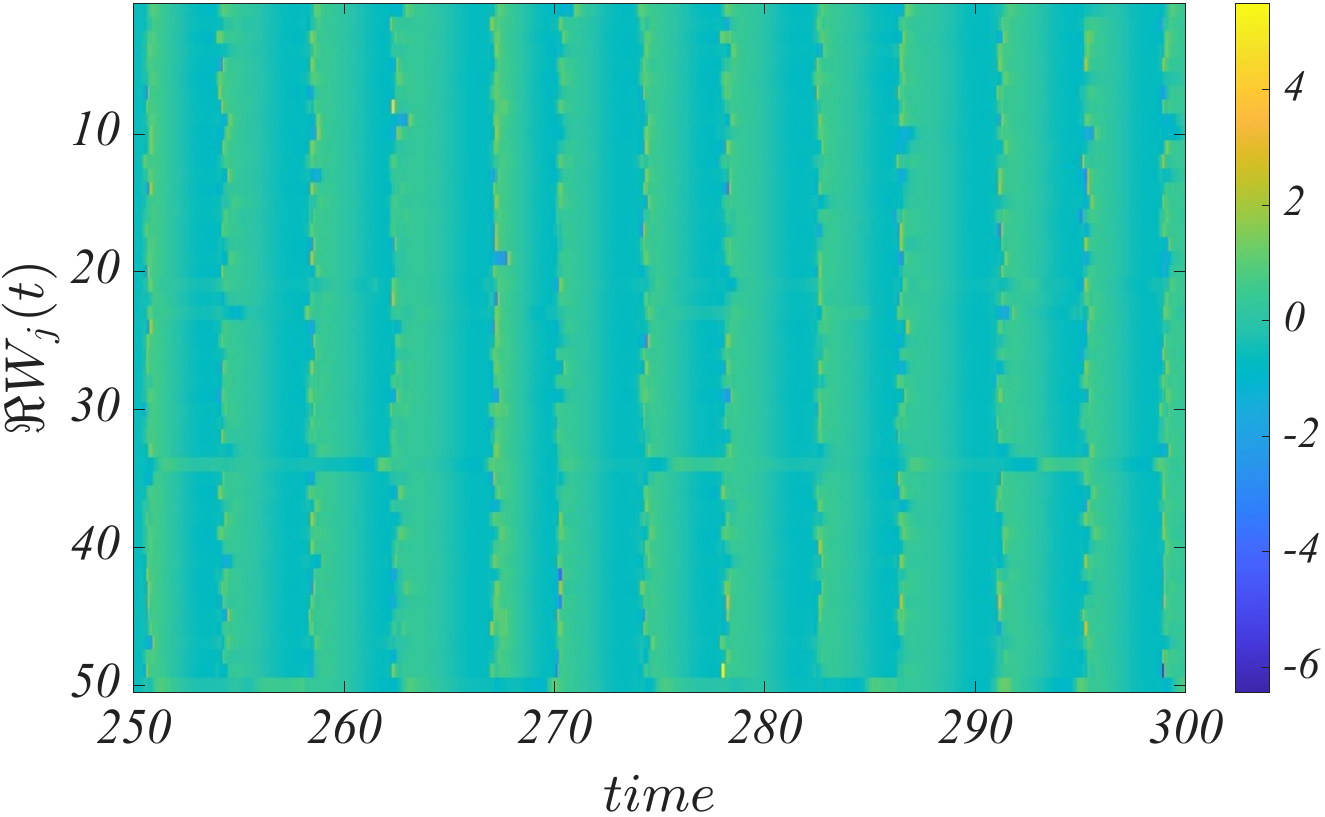}\quad     \includegraphics[width=0.4\linewidth]{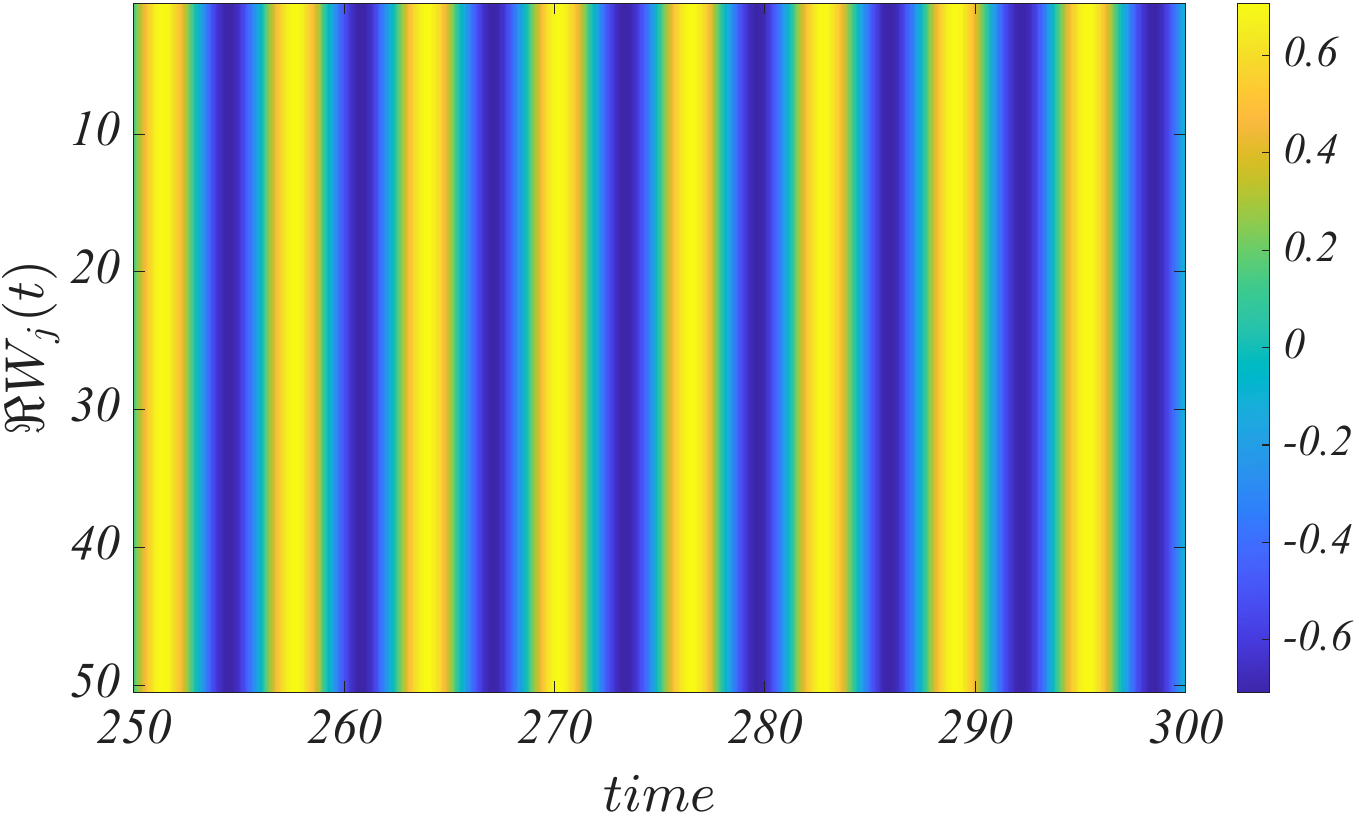}
    \caption{
    \textbf{Complete synchronization of Stuart-Landau oscillators coupled via a symmetric network, $a\neq b+1$.} The left column corresponds to the set of parameters $\sigma=0.5$, $\beta=1+2i$, $\mu=1+i$, $b=0$ and $a=2$, one can appreciate that the maximum Floquet exponent achieves positive values for some Laplace eigenvalue (red dot, top left panel) and the system cannot synchronize, as shown by the time evolution of $\Re W_j(t)$ (bottom left panel). The right column presents the results for $\sigma=0.5$, $\beta=1+2i$, $\mu=1+i$, $b=0$ and $a=5$. Now, the maximum Floquet exponent remains negative for all $\Lambda^{(\alpha)}$ (red dot, top right panel) and the system completely synchronizes (bottom right panel). In both top panels, the function $\zeta(x)$ is shown as eye-guide (blue curve). The underlying network is a random Erd\H{o}s-R\'enyi graph made of $N=50$ nodes and the probability to have a link is given by $p=0.1$.}
   \label{fig:NonautoUndirectednetwork}
\end{figure}

In general, exactly solving~\eqref{eq:Mode} is not possible and an explicit form for the Floquet exponents cannot be obtained. We can, however, turn to numerical integration to obtain a very good approximation of the latter. Let us now show that, by considering $a\neq b+1$, i.e., dealing with a periodic linearized system, can inhibit the emergence of synchronization for parameters allowing complete synchronization when $a=b+1$. In the left column of Fig.~\ref{fig:NonautoUndirectednetwork}, we consider the same model parameters used to obtain the results shown in the left column of Fig.~\ref{fig:AutoUnidrectednetwork}, i.e., $\sigma=0.5$, $\beta=1+2i$, $\mu=1+i$, $b=0$ but $a=2$. We can observe that the maximum Floquet exponent now reaches positive values for some eigenvalues of the Laplace matrix (top left panel). This pushes the system away from synchronization, as we can appreciate by looking at $\Re W_j(t)$ (bottom left panel). Hence, we can lose synchronization by slightly increasing the value of $a$. However, it is difficult to exactly determine the role of $a$. In fact, the results presented in the right column of Fig.~\ref{fig:NonautoUndirectednetwork} suggest that stronger nonlinearity, i.e., larger $a$, can favor the emergence of synchronization.

Let us observe that directionality can prevent the emergence of synchronization also in the case $a\neq b+1$. As for the former case, the reason for this is the presence of imaginary eigenvalues of the Laplace matrix. To support this claim, we consider the same setting used to obtain the results reported in the right column of Fig.~\ref{fig:NonautoUndirectednetwork}, namely, $\sigma=0.5$, $\beta=1+2i$, $\mu=1+i$, $b=0$ and $a=5$. However, now the support is a directed random Erd\H{o}s-R\'enyi graph made of $N=50$ nodes, with probability to have a link between two nodes given by $p=0.08$. We can observe that the maximum Floquet exponent is positive (left panel) or, equivalently, there are complex eigenvalues of the Laplace matrix falling into the instability region determined by $\zeta(\Lambda^{(\alpha)})>0$ (green region middle panel). Hence, the system cannot synchronize, as testified by the time evolution of $\Re W_j(t)$ (right panel).


\begin{figure}[h!]
    \centering 
       \includegraphics[width=0.3 \linewidth, height=2.8cm]{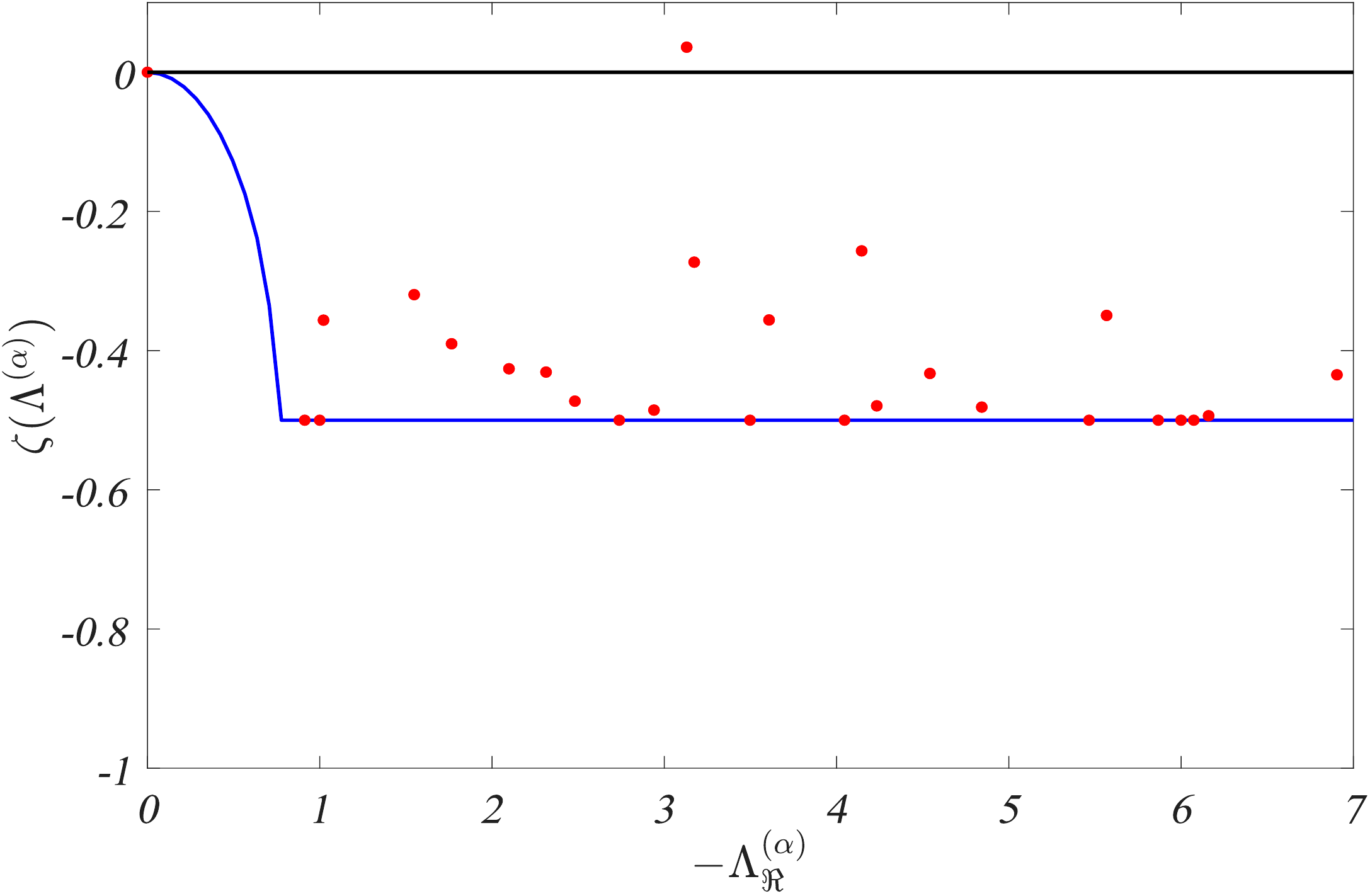}\quad  \includegraphics[width=0.3 \linewidth, height=2.8cm]{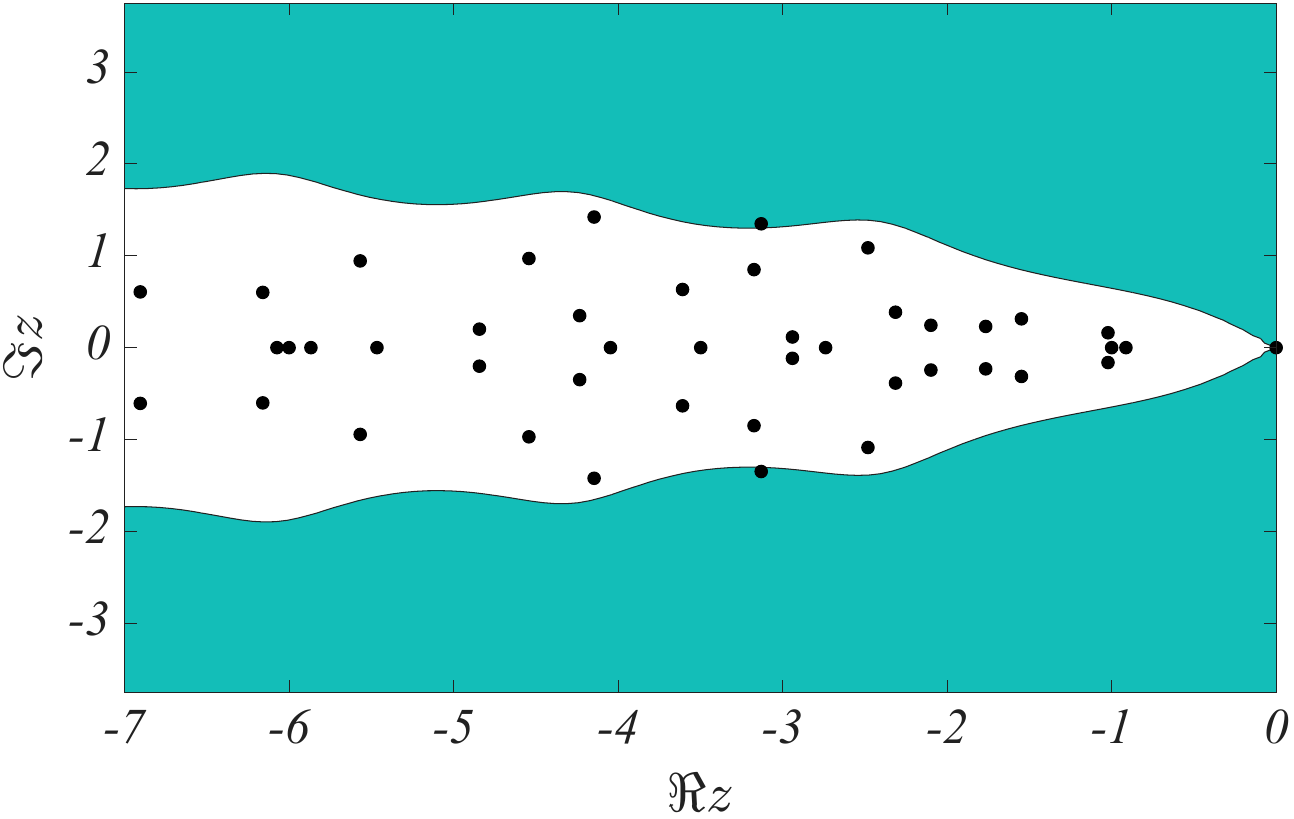} \quad       \includegraphics[width=0.3 \linewidth, height=2.8cm]{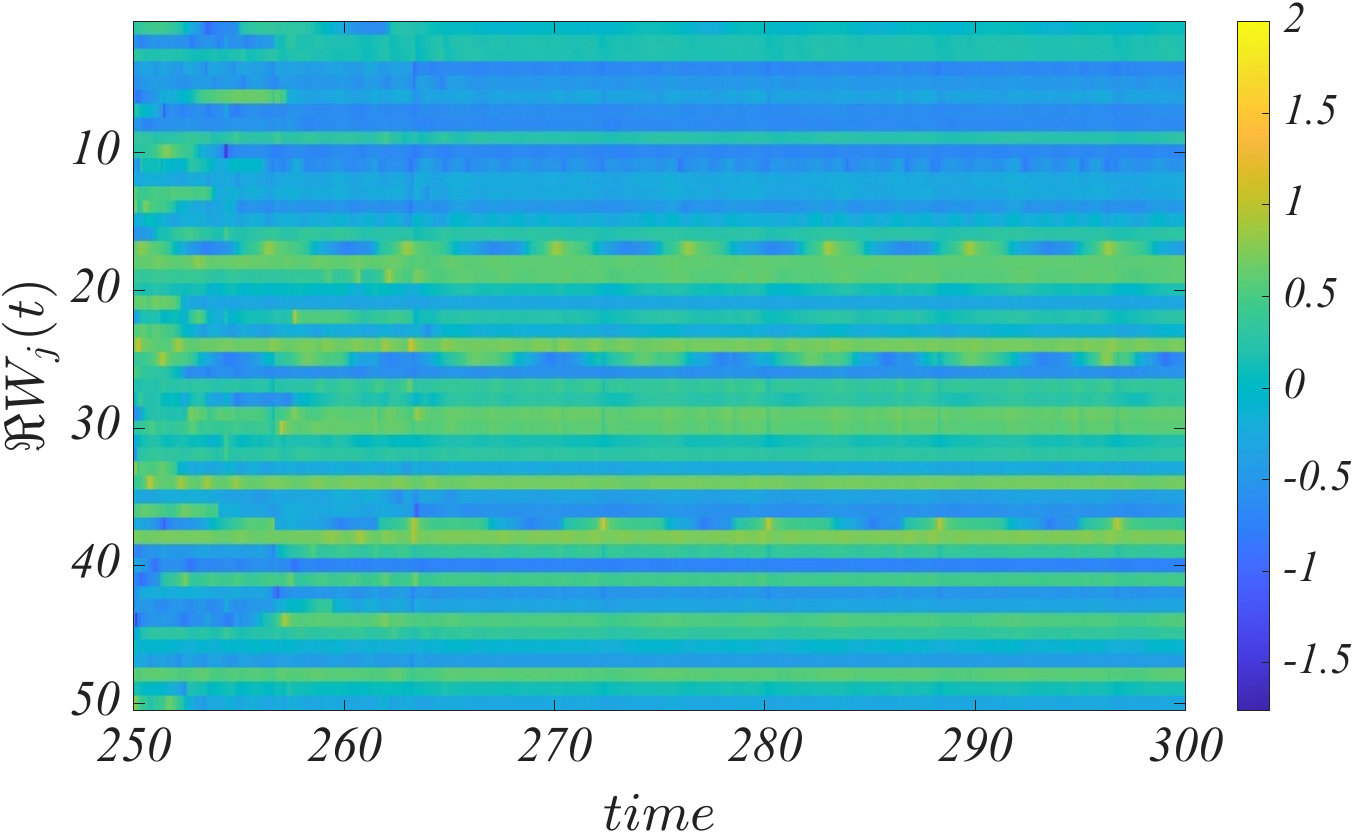}
    \caption{
    \textbf{Absence of complete synchronization of Stuart-Landau oscillators coupled via a directed network, $a\neq b+1$.} The left panel shows the maximum Floquet exponent (red dots), together with the function $\zeta(x)$ shown as eye-guide (blue curve). We can observe that $\zeta(\Lambda^{(\alpha)})$ takes a positive value in correspondence of a given $\Lambda^{(\alpha)}$. The middle panel provides a complementary view of the former panel, by reporting the region of instability in the complex plane, i.e., where the maximum Floquet exponent is positive. There, we can identify again a couple of complex conjugated eigenvalues lying in the instability region. The system can thus not synchronize, as shown in the right panel, where we plot $\Re W_j(t)$. The model parameters are $\sigma=0.5$, $\beta=1+2i$, $\mu=1+i$, $b=0$ and $a=5$; the underlying network is a directed random Erd\H{o}s-R\'enyi graph made of $N=50$ nodes, and the probability to have a link is given by $p=0.08$.}
   \label{fig:Nonautodirectednetworkab}
\end{figure}

\begin{figure}[h!]
    \centering 
\includegraphics[width=0.45\linewidth]{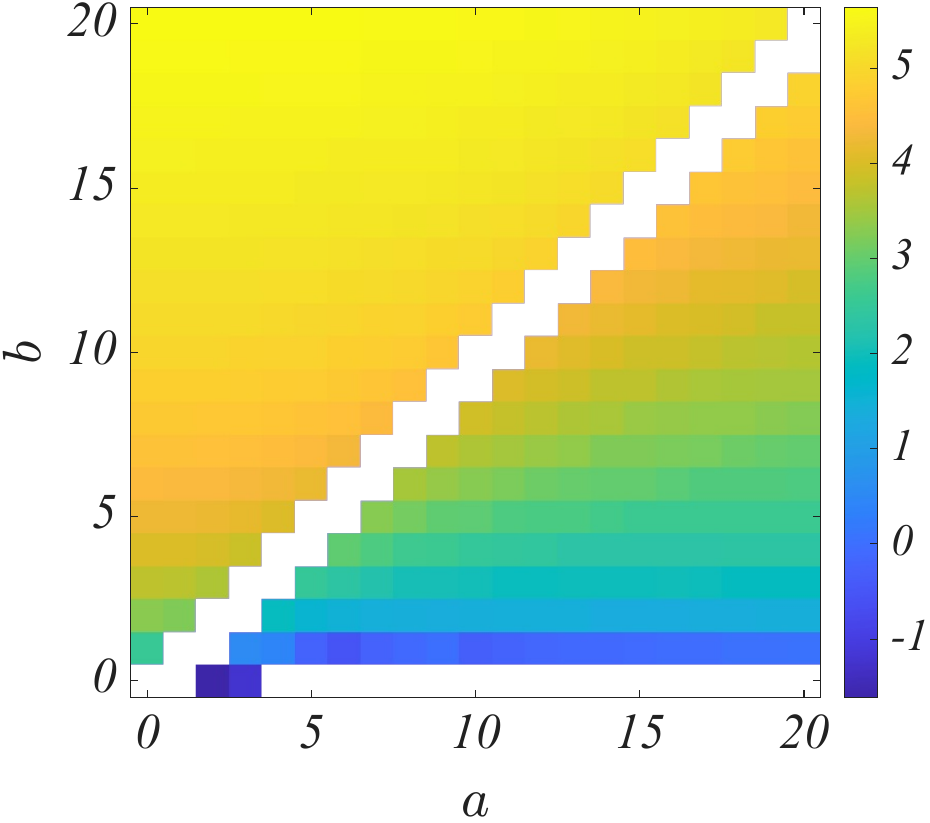}\quad\includegraphics[width=0.45\linewidth]{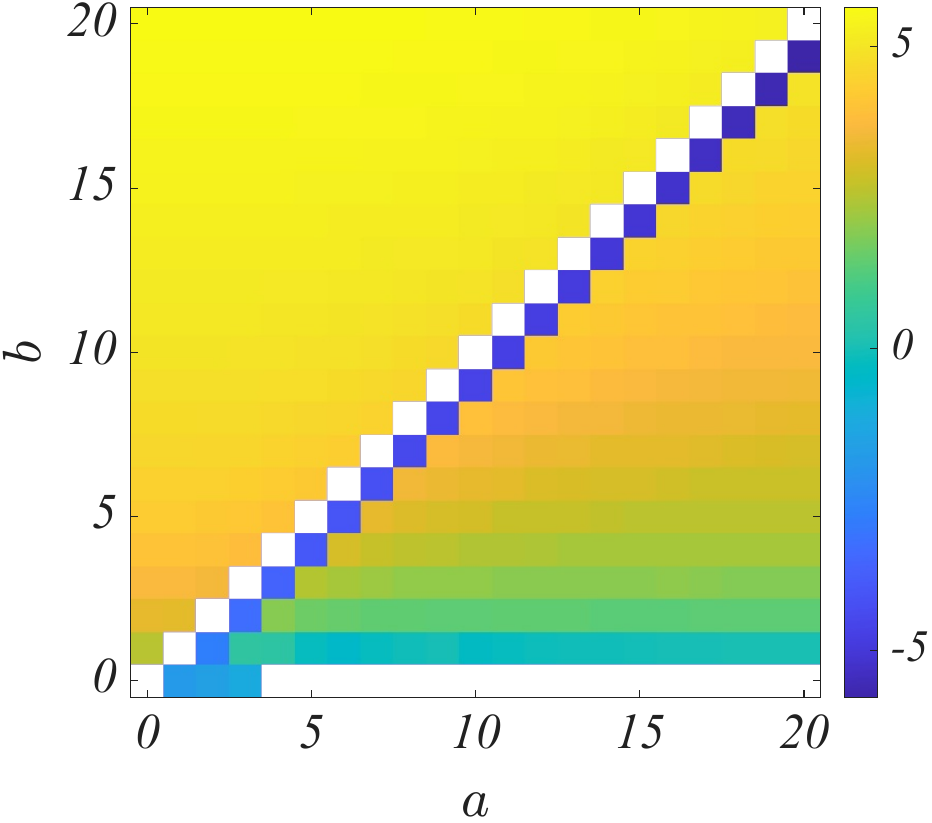}
    \caption{
    \textbf{Logarithm of the maximum Floquet exponent as a function of $a$ and $b$.} We report the logarithm of $\max_x\zeta (x)$ as a function of the exponents of the nonlinear coupling, $a$ and $b$, for the set of parameters $\sigma = 1$, $\beta = 1+2i$, $\mu=1+i$ (left panel) and $\sigma = 1$, $\beta = 1+2i$ and $\mu=1-i$ (right panel). Those values have been used the former in the left panel of Fig.~\ref{fig:AutoUnidrectednetwork} and the latter in the remaining panels of Fig.~\ref{fig:AutoUnidrectednetwork}. White squares correspond to $\max_x\zeta(x) \leq 0$ and thus its logarithm is not defined; system desynchronization is shown with squares with blue shades denoting $0<\max_x\zeta(x) <1$ and thus a negative $\log\left[\max_x\zeta(x)\right]$, while green-yellow square to $\max_x\zeta(x)>1$, hence $\log \max_x\zeta(x) >0$.}
   \label{fig:globalviewab}
\end{figure}

To achieve a better understanding of the interplay between the exponents $a$ and $b$ in the nonlinear coupling term, we numerically computed the largest Floquet exponent as a function of them for the remaining model parameters fixed to the values used to obtain the results of Fig.~\ref{fig:AutoUnidrectednetwork}. The results are reported in Fig.~\ref{fig:globalviewab}. Negative values of $\max_x\zeta(x)$, denoting the emergence of complete synchronization, are reported in white, while $0<\max_x\zeta(x) < 1$, associated to lack of synchronization, are shown with shades of blue, and $\max_x\zeta(x) > 1$, also representing lack of synchronization, are plotted in green-yellow. The left panel corresponds to parameters $\sigma = 1$, $\beta = 1+2i$, $\mu=1+i$ (same as for the left panel of Fig.~\ref{fig:AutoUnidrectednetwork}) while $\sigma = 1$, $\beta = 1+2i$, $\mu=1-i$ in the right panel (same as for the middle and right panels of Fig.~\ref{fig:AutoUnidrectednetwork}). As observed in the left panel of Fig.~\ref{fig:AutoUnidrectednetwork}, for such set of parameters the system synchronizes in the autonomous case, i.e., $a=b+1$: indeed, we observe the same phenomenon for all $(a,b)$ such that $a=b+1$ (see white squares on the upper diagonal of the left panel of Fig.~\ref{fig:globalviewab}). On the other hand the middle and right panels of Fig.~\ref{fig:AutoUnidrectednetwork} show that the system cannot synchronize for such set of parameters in the autonomous case, and this holds true for all $a=b+1$ (see blue squares on the upper diagonal of the right panel of Fig.~\ref{fig:globalviewab}). We can thus conclude that, with the theory presented in Section~\ref{ssec:autonom}, we can completely describe the region of parameters for which $a=b+1$. The remaining regions can be explained by resorting to the Floquet analysis, as shown in Section~\ref{ssec:nonauton}. Finally, in both panels of Fig.~\ref{fig:globalviewab}, we can observe the presence of non positive values of $\log \max_x\zeta(x)$ (white squares) for $a=b$ and $a\geq 4$ and $b=0$. Let us hereby prove the former claim ($a=b$), while the latter will be discussed in Section~\ref{sec:approxFloquet}.

Let us thus assume $a=b$. Hence, Eq.~\eqref{eq:SLLinearmodelGraphTimDep} simplifies into
\begin{equation} 
\label{eq:SLLinearmodelGraphaeqb}
\frac{d}{dt}    \left(\begin{array}{cc}
        {\hat{\rho}}_{\alpha}   \\
        {\hat{\theta}}_{\alpha} 
    \end{array}\right) =
    \left[\left(\begin{array}{cc}
        -2\sigma_{\Re} & 0 \\
         -2\beta_{\Im}\dfrac{\sigma_{\Re}}{\beta_{\Re}}&  0
    \end{array}\right)+2a|W_{\mathrm{LC}}|^{2a-1}\Lambda^{(\alpha)}\left(\begin{array}{cc}
         T_{1}(t) & 0 \\
         T_{2}(t) & 0 
    \end{array}\right)\right]\left(\begin{array}{cc}
        \hat{\rho}_{\alpha}   \\
        \hat{\theta}_{\alpha} 
    \end{array}\right) 
\end{equation}
with
\begin{equation*}
 T_1(t)=\mu_{\Re} \cos(\omega t) + \mu_{\Im} \sin(\omega t) \text{ and }T_2(t)=\mu_{\Im} \cos(\omega t) - \mu_{\Re} \sin(\omega t)\, .
\end{equation*}
The equation ruling the evolution of $\hat{\rho}_{\alpha}$ is then given by
\begin{equation*} 
        \frac{1}{\hat{\rho}_{\alpha}}\frac{d\hat{\rho}_{\alpha}}{dt}= \frac{d }{dt}\log(\hat{\rho}_{\alpha}) =  -2\sigma_{\Re} +2a|W_{\mathrm{LC}}|^{2a-1}\Lambda^{(\alpha)}
         [\mu_{\Re} \cos(\omega t) + \mu_{\Im} \sin(\omega t)]\, ,
\end{equation*}
which can be explicitly solved to obtain
\begin{equation}
    \hat{\rho}_{\alpha}(t)=\hat{\rho}_{\alpha}(0) \exp \left( -2\sigma_{\Re}t +2|W_{LC}|^{2a-1}\Lambda^{(\alpha)}
         \dfrac{a}{\omega}[\mu_{\Re} \sin(\omega t) - \mu_{\Im} \cos(\omega t)+\mu_{\Im}] \right)\, .
         \label{eq:rho_solution}
\end{equation}
We can insert this result into the differential equation for $\hat{\theta}_{\alpha}$ and get
\begin{equation}
\label{eq:dotthetaaeqb}
\begin{aligned}
\frac{d\hat{\theta}_{\alpha}}{dt} 
&= \hat{\rho}_{\alpha}(0)
\Bigl[-2\beta_\Im \frac{\sigma_\Re}{\beta_\Re}
    + 2a |W_{\mathrm{LC}}|^{2a-1}\Lambda^{(\alpha)}T_2(t)\Bigr] \\
&\quad \times
\exp\!\left[-2\sigma_{\Re}t 
    + 2|W_{LC}|^{2a-1}\Lambda^{(\alpha)}\frac{a}{\omega}
    (\mu_{\Re} \sin(\omega t) - \mu_{\Im} \cos(\omega t) + \mu_{\Im})
\right] .
\end{aligned}
\end{equation}
From the above expression, we can conclude that $\hat{\theta}_{\alpha}(t)$ is given by
\begin{equation*}
\hat{\theta}_{\alpha}(t) = \hat{\theta}_{\alpha}(0)+\hat{\rho}_\alpha(0)\Theta(t)\, ,
\end{equation*}
where $\Theta(t)$ is the integral of the right hand side of~\eqref{eq:dotthetaaeqb}. Note that we have emphasized the role of $\hat{\rho}_\alpha(0)$. This means that we can exactly solve Eq.~\eqref{eq:Mode} to obtain the monodromy matrix
\begin{equation}
\label{eq:monodmataeqb}
\mathbf{C}_{\alpha}= \mathbf{M}_{\alpha}(T)=\left(
\begin{matrix}
 e^{-2\sigma_\Re T} & 0 \\
 \Theta(T) & 1
\end{matrix}
\right)\, ,
\end{equation}
and then compute the Floquet exponents
\begin{equation*}
 \zeta_{\alpha,1}=-2\sigma_\Re \text{ and } \zeta_{\alpha,2}=0\, .
\end{equation*}
By exploiting the triangular shape of the monodromy matrix~\eqref{eq:monodmataeqb}, we can write, for all positive integer $k$,
\begin{equation*}
 \left(
\begin{matrix}
 \hat{\rho}_\alpha(kT)\\
\hat{\theta}_\alpha(kT)
\end{matrix}
\right)= \left(
\begin{matrix}
 e^{-2\sigma_\Re kT} \hat{\rho}_\alpha(0)\\
\hat{\theta}_\alpha(0)+ \Theta(T) \hat{\rho}_\alpha(0)\frac{1-e^{-2\sigma_\Re (k+1)T}}{1-e^{-2\sigma_\Re T}}\\
\end{matrix}
\right)\, .
\end{equation*}
Going back to the original variables, we obtain~\eqref{eq:projeigenb}
\begin{equation*}
\begin{aligned}
\begin{pmatrix}
\rho_j(kT)\\
\theta_j(kT)
\end{pmatrix}
&= \sum_\alpha \Psi_j^{(\alpha)}
\begin{pmatrix}
\hat{\rho}_\alpha(kT)\\
\hat{\theta}_\alpha(kT)
\end{pmatrix} = \sum_\alpha \Psi_j^{(\alpha)}
\begin{pmatrix}
e^{-2\sigma_\Re kT} \hat{\rho}_\alpha(0)\\
\hat{\theta}_\alpha(0)+ \Theta(T)\hat{\rho}_\alpha(0)
\dfrac{1-e^{-2\sigma_\Re (k+1)T}}{1-e^{-2\sigma_\Re T}}
\end{pmatrix} \\
&= 
\begin{pmatrix}
\rho_j(0)e^{-2\sigma_\Re kT}\\
\theta_j(0)+\rho_j(0)\Theta(T)
\dfrac{1-e^{-2\sigma_\Re (k+1)T}}{1-e^{-2\sigma_\Re T}}
\end{pmatrix}\, .
\end{aligned}
\end{equation*}
We can eventually conclude that, for all $j=1,\dots,N$ and $k$ large enough,
\begin{equation}
\label{eq:concaeqb}
\rho_j(kT)\rightarrow 0 \text{ and }\theta_j(kT)\rightarrow\theta_j(0)+\rho_j(0)\Theta(T) \frac{1}{1-e^{-2\sigma_\Re T}}\, ,
\end{equation}
hence, the angles will be phase-locked (in the linear approximation). Indeed, we have that 
\begin{equation*}
\theta_j(kT)-\theta_\ell(kT) \rightarrow \theta_j(0)-\theta_\ell(0)+\left(\rho_j(0)-\rho_\ell(0)\right)\Theta(T)\frac{1}{1-e^{-2\sigma_\Re T}}\, .
\end{equation*}
Let us observe that this result is independent of the network topology and is solely governed by intrinsic parameters. Let us also stress that the full nonlinear system could exhibit a different behavior, this case being degenerate: in fact, the maximum Floquet exponent vanishes for all values of $\Lambda^{(\alpha)}$, meaning that all modes are neutrally stable.

\section{Analytical approximation of the Floquet exponent}
\label{sec:approxFloquet}

In the previous Sections, we have shown that, for the case of $a \neq b+1$, the stability of the synchronous solution, i.e., the emergence of complete synchronization, can be assessed by studying the non-autonomous linearized system~\eqref{eq:SLLinearmodelGraphTimDep}. This being periodic, we can thus rely on Floquet theory. If the largest real part of the Floquet exponent is positive, then the system cannot synchronize. The aim of this Section is to provide a (semi) analytical approximation of the Floquet exponent grounded on the use of the Jacobi-Anger expansion~\cite{AS1964} of sine and cosine functions in terms of Bessel functions of first kind, $J_n(z)$:
\begin{eqnarray}
\label{eq:JAexp}
\displaystyle \cos(z\cos(\theta)) &=& J_0(z)+2\sum_{n=1}^\infty (-1)^n J_{2n}(z)\cos(2n\theta)\, ,\\
\displaystyle \sin(z\cos(\theta)) &=& -2\sum_{n=1}^\infty (-1)^n J_{2n-1}(z)\cos[(2n-1)\theta]\notag\, ,\\
\displaystyle \cos(z\sin(\theta)) &=& J_0(z)+2\sum_{n=1}^\infty  J_{2n}(z)\cos(2n\theta)\notag\, ,\\
\displaystyle \sin(z\sin(\theta)) &=& 2\sum_{n=1}^\infty (-1)^n J_{2n-1}(z)\cos[(2n-1)\theta]\notag\, ,
\end{eqnarray}
In particular, we will consider $b=0$ and $a\geq 2$ to provide an explanation of the results reported in Fig.~\ref{fig:globalviewab} by assuming the same values of the parameters. Let us rewrite the linearized system~\eqref{eq:SLLinearmodelGraphTimDep} in the following form, where, to lighten the notation, we used generic variables $x$ and $y$ and dropped the indexes $\alpha$:
\begin{equation}
\label{eq:STLinearmodelGraph1matB}
\frac{d}{dt}\left(
\begin{matrix}
 x\\y
\end{matrix}
\right)=\left(\begin{matrix}
a_{11} & a_{12}\\a_{21} & a_{22}
\end{matrix}\right)\left(
\begin{matrix}
 x\\y
\end{matrix}
\right)\, ,
\end{equation}
with
\begin{equation}
\label{eq:ai}
\begin{aligned}
a_{11}(t) &= -2\sigma_{\Re} + a|W_{\mathrm{LC}}|^{a-1}\Lambda |\mu| \cos(\Omega t + \arg \mu), \quad
a_{12}(t) = -a|W_{\mathrm{LC}}|^{a-1}\Lambda |\mu| \sin(\Omega t + \arg \mu),\\
a_{21}(t) &= -2\beta_{\Im} \frac{\sigma_{\Re}}{\beta_\Re} + a|W_{\mathrm{LC}}|^{a-1}\Lambda |\mu| \sin(\Omega t + \arg \mu), \quad
a_{22}(t) = a|W_{\mathrm{LC}}|^{a-1}\Lambda |\mu| \cos(\Omega t + \arg \mu)\, ,
\end{aligned}
\end{equation}
where we rewrote $\mu=|\mu|e^{i \arg\mu}$ and defined $\Omega=(a-1)\omega$.

To proceed further, let us introduce a change of variable to polar coordinates
\begin{equation}
\label{eq:polcoordB}
x=r\cos\phi \text{ and }y=r\sin\phi\, ,
\end{equation}
and obtain the time evolution of $r$ and $\phi$ from Eq.~\eqref{eq:STLinearmodelGraph1matB}. A simple computation allows to write
\begin{equation}
\label{eq:dotrdotthetaB}
\begin{cases}
 \dfrac{1}{r}\dfrac{dr}{dt} & = \dfrac{a_{11}+a_{22}}{2}+\dfrac{a_{11}-a_{22}}{2}\cos(2\phi)+\dfrac{a_{12}+a_{21}}{2}\sin(2\phi)\\
 \dfrac{d\phi}{dt}&=\dfrac{a_{21}-a_{12}}{2}+\dfrac{a_{21}+a_{12}}{2}\cos(2\phi)+\dfrac{a_{22}-a_{11}}{2}\sin(2\phi)\, .
\end{cases}
\end{equation}
Before proceeding, let us observe that the equation for $d\phi/ dt$ does not depend on the variable $r$. The time evolution equation of $\phi$ is thus a scalar non-autonomous ODE. Once the solution has been found, it can be substituted into the first equation and solved through direct integration, obtaining thus $\log[r(t)/r(0)]$. 

The coefficients involved in~\eqref{eq:dotrdotthetaB} can be obtained from~\eqref{eq:ai}, namely,
\begin{eqnarray}
\label{eq:aib0a1B}
\frac{a_{11}+a_{22}}{2}&=&-\sigma_{\Re} +a|W_{\mathrm{LC}}|^{a-1}\Lambda |\mu| \cos(\Omega t +\arg \mu)  \, ,\quad \frac{a_{11}-a_{22}}{2}=-\sigma_{\Re}\notag \, ,\\
\frac{a_{12}+a_{12}}{2}&=&-\beta_{\Im}\frac{\sigma_{\Re}}{\beta_\Re} \, ,\quad \frac{a_{21}-a_{12}}{2}=-\beta_{\Im}\frac{\sigma_{\Re}}{\beta_\Re}+a |W_{\mathrm{LC}}|^{a-1}\Lambda |\mu| \sin(\Omega t +\arg \mu) \, ,
\end{eqnarray}
 and, thus, Eq.~\eqref{eq:dotrdotthetaB} rewrites as
\begin{equation}
\label{eq:timeevolrphi}
\left\{
\begin{aligned}
 \dfrac{1}{r}\dfrac{dr}{dt} &= -\sigma_{\Re}-\sigma_{\Re} \cos(2\phi)
 -\beta_{\Im}\frac{\sigma_{\Re}}{\beta_\Re}\sin(2\phi)
 +a|W_{\mathrm{LC}}|^{a-1}\Lambda |\mu| \cos(\Omega t+\arg\mu) \\
 \dfrac{d\phi}{dt} &= -\beta_{\Im}\frac{\sigma_{\Re}}{\beta_\Re}
 -\beta_{\Im}\frac{\sigma_{\Re}}{\beta_\Re}\cos(2\phi)
 +\sigma_{\Re}\sin(2\phi)
 +a|W_{\mathrm{LC}}|^{a-1}\Lambda |\mu| \sin(\Omega t+\arg\mu)\, .
\end{aligned}
\right.
\end{equation}
To solve the equation for $\phi$, let us make use of the following ansatz
\begin{equation}
\label{eq:thetaanss}
\phi(t) =-k \frac{\Omega t}{2}+A_k+B_k e^{i\Omega t}+\bar{B}_ke^{-i\Omega t}=-k \frac{\Omega t}{2}+A_k+2|B_k|\cos(\Omega t +\arg B_k)\, ,
\end{equation}
with $k\in \mathbf{N}\cup \{0\}$. Namely, except the possible linear trend given by $-k \Omega t/2$, the function oscillates with frequency $\Omega$, hereby described by the first Fourier term, with $A_k\in\mathbb{R}$ and $B_k=|B_k|e^{i\arg B_k}\in\mathbb{C}$. We are thus looking for a $1$-parameter family of solutions $\phi(t)$ indexed by the integer $k\geq 0$, whose time evolution is, in first approximation, linear and given by $-k\Omega t/2$, upon which an oscillatory behavior is superposed.

To proceed, we insert the ansatz~\eqref{eq:thetaanss} into the second equation of~\eqref{eq:timeevolrphi} and, equating same Fourier modes, we determine in this way the parameters $A_k$ and $B_k$ for all $k$. Then, with this information, we solve the first equation of~\eqref{eq:timeevolrphi} by direct integration. Let us observe that the determination of the Fourier coefficients and the integration will be performed numerically, because the involved integrals cannot computed explicitly. This is the reason why we called the method semi-analytical. 

By inserting~\eqref{eq:thetaanss} into the second equation of~\eqref{eq:timeevolrphi}, we get
\begin{eqnarray}
\label{eq:derthetaanss}
\frac{d\phi}{dt} &=&-k \frac{\Omega}{2}-2\Omega |B_k|\sin(\Omega t +\arg B_k) \\
&=&-\beta_{\Im}\frac{\sigma_{\Re}}{\beta_\Re}-\beta_{\Im}\frac{\sigma_{\Re}}{\beta_\Re}\cos(2\phi)+\sigma_{\Re}\sin(2\phi)+ a|W_{\mathrm{LC}}|^{a-1}\Lambda |\mu| \sin(\Omega t+\arg\mu)\notag\\
&=&-\beta_{\Im}\frac{\sigma_{\Re}}{\beta_\Re}-\beta_{\Im}\frac{\sigma_{\Re}}{\beta_\Re}\cos\left(-k \Omega t+2A_k+4|B_k|\cos(\Omega t +\arg B_k)\right)+ \notag\\
&+& \sigma_{\Re}\sin\left(-k \Omega t+2A_k+4|B_k|\cos(\Omega t +\arg B_k)\right)+\notag a|W_{\mathrm{LC}}|^{a-1}\Lambda |\mu| \sin(\Omega t+\arg\mu)\, ,
\end{eqnarray}

Let us compute $\cos(2\phi)$ and $\sin(2\phi)$, with $\phi$ given by Eq.~\eqref{eq:thetaanss} for a generic $k$, by using the Jacobi-Anger expressions~\eqref{eq:JAexp} and some trigonometry to finally get
\begin{eqnarray}
\label{eq:cos2phi}
\cos(2\phi) &=& J_0(4|B_k|) \cos(-k\Omega t + 2A_k) \notag\\
&+& \sum_{n \geq 1} (-1)^n J_{2n}(4|B_k|) 
\Big[ \cos\big( (2n - k)\Omega t + 2A_k + 2n \arg B_k \big) \notag\\
&+& \cos\big( (2n + k)\Omega t - 2A_k + 2n \arg B_k \big) \Big] \notag\\
&+& \sum_{n \geq 1} (-1)^n J_{2n-1}(4|B_k|) 
\Big[ \sin\big( (2n - 1 - k)\Omega t + 2A_k + (2n - 1)\arg B_k \big) \notag\\
&-& \sin\big( (2n - 1 + k)\Omega t - 2A_k + (2n - 1)\arg B_k \big) \Big] \, ,
\end{eqnarray}
and
\begin{eqnarray}
\label{eq:sin2phi}
\sin(2\phi) &=& J_0(4|B_k|) \sin(-k\Omega t + 2A_k) \notag\\
&+& \sum_{n \geq 1} (-1)^n J_{2n}(4|B_k|) 
\Big[ \sin\big( (2n - k)\Omega t + 2A_k + 2n \arg B_k \big) \notag\\
&-& \sin\big( (2n + k)\Omega t - 2A_k + 2n \arg B_k \big) \Big] \notag\\
&-& \sum_{n \geq 1} (-1)^n J_{2n-1}(4|B_k|) 
\Big[ \cos\big( (2n - 1 - k)\Omega t + 2A_k + (2n - 1)\arg B_k \big) \notag\\
&+& \cos\big( (2n - 1 + k)\Omega t - 2A_k + (2n - 1)\arg B_k \big) \Big] \, .
\end{eqnarray}
We now want to determine the Fourier expansion of both terms in the equality~\eqref{eq:derthetaanss} and equate the modes associated to the same harmonics. Let us then fix $k=0$ in the ansatz~\eqref{eq:thetaanss}. Then, the $0$-th mode can be easily found by looking at Eqs.~\eqref{eq:cos2phi} and~\eqref{eq:sin2phi}, from whichwe obtain
\begin{equation}
\label{eq:avecossin}
\langle \cos(2\phi)\rangle_{0} = J_0(4|B_0|)\cos(2A_0) \text{ and }\langle \sin(2\phi)\rangle_{0} = J_0(4|B_0|)\sin(2A_0)\, .
\end{equation}
From equating the $0$-th order modes on the left and right sides of Eq.~\eqref{eq:derthetaanss} for $k=0$, we obtain
\begin{equation}
\label{eq:avcosk0B}
0=-\beta_\Im \frac{\sigma_\Re}{\beta_\Re}\left(1+J_0(4|B_0|)\cos(2A_0)\right)+\sigma_\Re J_0(4|B_0|)\sin(2A_0)\, .
\end{equation}

Let us now compute the terms with $\sin(\Omega t)$ and $\cos(\Omega t)$, again for $k=0$ in~\eqref{eq:derthetaanss}. In the case of $\cos(2\phi)$, we get
\begin{equation*}
 \langle \cos(2\phi)\rangle_{1}=-J_{1}(4|B_0|)\left[\sin(\Omega t+2A_0+\arg B_0)-\sin(\Omega t -2A_0+\arg B_0)\right]\, ,
\end{equation*}
while, for $\sin(2\phi)$, we have 
\begin{equation*}
 \langle \sin(2\phi)\rangle_{1}=J_1(4|B_0|)\left[\cos(\Omega t +2A_0+\arg B0)+\cos(\Omega t -2A_0+\arg B_0)\right]\, .
\end{equation*}
Thus, $1$-st order modes, i.e., the coefficients of $\sin(\Omega t)$ and $\cos(\Omega t)$, in the left and right sides of Eq.~\eqref{eq:derthetaanss} for $k=0$, return
\begin{align*}
\label{eq:Omegatcosk0}
-2|B_0|\Omega \sin(\Omega t + \arg B_0)
&= \beta_\Im \frac{\sigma_\Re}{\beta_\Re} 
   J_{1}(4|B_0|)
   \big[\sin(\Omega t + 2A_0 + \arg B_0)
     - \sin(\Omega t - 2A_0 + \arg B_0)\big] \\
&\quad + \sigma_\Re J_1(4|B_0|)
   \big[\cos(\Omega t + 2A_0 + \arg B_0)
     + \cos(\Omega t - 2A_0 + \arg B_0)\big] \\
&\quad + a|W_{\mathrm{LC}}|^{a-1}\Lambda |\mu| 
   \sin(\Omega t + \arg\mu)\, .
\end{align*}

Now, by reorganizing the terms, we obtain
\begin{align*}
-2|B_0|\Omega \cos(\arg B_0)
&= \beta_\Im \frac{\sigma_\Re}{\beta_\Re}
    J_{1}(4|B_0|)
    \big[\cos(2A_0+\arg B_0)-\cos(-2A_0+\arg B_0)\big] \\
&\quad + \sigma_\Re J_1(4|B_0|)
    \big[-\sin(2A_0+\arg B_0)-\sin(-2A_0+\arg B_0)\big] \\
&\quad + a|W_{\mathrm{LC}}|^{a-1}\Lambda |\mu| \cos(\arg \mu), \\
-2|B_0|\Omega \sin(\arg B_0)
&= \beta_\Im \frac{\sigma_\Re}{\beta_\Re}
    J_{1}(4|B_0|)
    \big[\sin(2A_0+\arg B_0)-\sin(-2A_0+\arg B_0)\big] \\
&\quad + \beta' J_1(4|B_0|)
    \big[\cos(2A_0+\arg B_0)+\cos(-2A_0+\arg B_0)\big] \\
&\quad + a|W_{\mathrm{LC}}|^{a-1}\Lambda |\mu| \sin(\arg \mu)\, .
\end{align*}

In conclusion, we have to solve for $A_0$, $|B_0|$ and $\arg B_0$ the following system:
\begin{equation}
\label{eq:k0finalB}
\left\{
\begin{array}{rcl}
\beta_\Im \frac{\sigma_\Re}{\beta_\Re} &=& (-\beta_\Im \frac{\sigma_\Re}{\beta_\Re} \cos(2A_0)+\sigma_\Re \sin(2A_0))J_0(4|B_0|),\\
-2|B_0|\Omega \cos(\arg B_0) &=& -2J_1(4|B_0|)(\beta_\Im \frac{\sigma_\Re}{\beta_\Re} \sin(2A_0)\sin(\arg B_0)+\sigma_\Re \cos(2A_0)\sin(\arg B_0))\\
&& + a|W_{\mathrm{LC}}|^{a-1}\Lambda |\mu| \cos(\arg \mu),\\
-2|B_0|\Omega \sin(\arg B_0) &=& 2J_1(4|B_0|)(\beta_\Im \frac{\sigma_\Re}{\beta_\Re} \sin(2A_0)\cos(\arg B_0)+\sigma_\Re \cos(2A_0)\cos(\arg B_0))\\
&& + a|W_{\mathrm{LC}}|^{a-1}\Lambda |\mu| \sin(\arg \mu)\, .
\end{array}
\right.
\end{equation}

Once $A_0$, $|B_0|$ and $\arg B_0$ have been obtained, we can use this information to determine the growth rate of $dr/dt$ given by~\eqref{eq:timeevolrphi}. By integrating the latter equation from $t=0$ to $t=qT$ for some integer $q$ (recall that $T=2\pi/\Omega$), we realize that the only terms in the right-hand side that do not vanish are those associated to the average. Hence,
\begin{equation*}
 \int_0^{qT} \frac{1}{r}\frac{dr}{dt} \,dt= -\sigma_\Re qT-\sigma_\Re  \int_0^{qT}\cos(2\phi(t))\, dt-\beta_\Im \frac{\sigma_\Re}{\beta_\Re} \int_0^{qT}\sin(2\phi(t))\, dt\, ,
\end{equation*}
from which, recalling~\eqref{eq:avecossin}, we get
\begin{equation*}
\log\frac{r(qT)}{r(0)}=qT\left[-\sigma_\Re -J_0(4|B_0|) \left(\sigma_\Re  \cos(2A_0)+\beta_\Im \frac{\sigma_\Re}{\beta_\Re} \sin(2A_0)\right)\right]\, .
\end{equation*}
Hence, the stability of $r$ is obtained by the condition
\begin{equation}
\label{eq:k0stabr}
\Sigma_0(\Lambda)=-\sigma_\Re\left[1+J_0(4|B_0|) \left( \cos(2A_0)+ \frac{\beta_\Im}{\beta_\Re}  \sin(2A_0)\right)\right]<0\, ,
\end{equation}
where we emphasized the dependence on the eigenvalue of the Laplace matrix $\Lambda$, ``hidden'' in the reconstructed variables $A_0$, $|B_0|$ and $\arg B_0$. 

We also used the index $\Sigma_0$ to recall that the computation has been done by assuming $k=0$ in Eq.~\eqref{eq:thetaanss}. Let us observe that a similar computation can be performed of all $k\geq 0$. Indeed, for $k=1$, we can show that the unknown $A_1$, $|B_1|$ and $\arg B_1$ are solutions of the following system, where $A_0$, $|B_0|$ and $\arg B_0$ have been determined in the previous step:
\begin{equation}
\label{eq:k1final}
\left\{
\begin{array}{rcl}
-\dfrac{\Omega}{2} &=& 
 -\beta_\Im \dfrac{\sigma_\Re}{\beta_\Re} 
 \\
 && + (\beta_\Im \dfrac{\sigma_\Re}{\beta_\Re} \sin(2A_1+\arg B_1) + \sigma_\Re \cos(2A_0+\arg B_1)) J_1(4|B_1|),\\
-2|B_1|\Omega \cos(\arg B_1) &=&
 -J_{0}(4|B_1|)(\beta_\Im \dfrac{\sigma_\Re}{\beta_\Re} \sin(2A_1)+\sigma_\Re \cos(2A_1))\\
&& - J_2(4|B_1|)(\beta_\Im \dfrac{\sigma_\Re}{\beta_\Re} \sin(2A_1+2\arg B_1)+\sigma_\Re \cos(2A_1+2\arg B_1))\\
&& + a|W_{\mathrm{LC}}|^{a-1}\Lambda |\mu| \cos(\arg \mu),\\
-2|B_1|\Omega \sin(\arg B_1) &=&
 -J_{0}(4|B_1|)(\beta_\Im \dfrac{\sigma_\Re}{\beta_\Re} \cos(2A_1)-\sigma_\Re \sin(2A_1))\\
&& - J_2(4|B_1|)(-\beta_\Im \dfrac{\sigma_\Re}{\beta_\Re} \cos(2A_1+2\arg B_1)+\sigma_\Re \sin(2A_1+2\arg B_1))\\
&& + a|W_{\mathrm{LC}}|^{a-1}\Lambda |\mu| \sin(\arg \mu)\, .
\end{array}
\right.
\end{equation}

By integrating again the first equation of~\eqref{eq:timeevolrphi}, we obtain the time evolution of $\log r(t)/r(0)$. Then, the stability is given by the condition
\begin{equation}
\label{eq:k1stabr}
\Sigma_1(\Lambda)=-\sigma_\Re\left[1 -J_1(4|B_1|) \left(  \sin(2A_1+\arg B_1)- \frac{\beta_\Im}{\beta_\Re}  \cos(2A_1+\arg B_1)\right)\right]<0\, ,
\end{equation}
where we emphasized again the dependence on $\Lambda$ via the coefficients $A_1$ and $B_1$.

\begin{figure}[h!]
    \centering 
    \includegraphics[width=\textwidth]{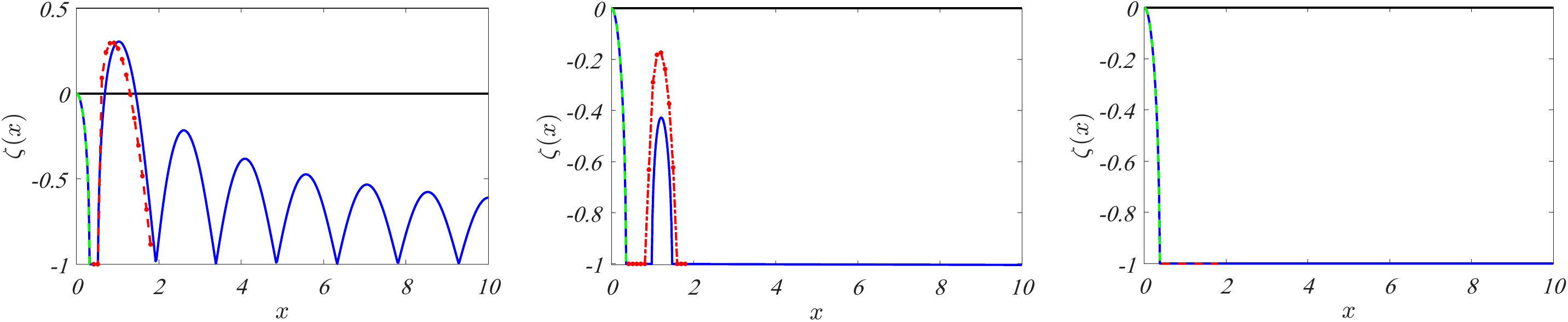}\quad    
    \caption{
    \textbf{Semi-analytical approximation of the maximal Floquet exponent.} For the choice of parameters $\sigma = 1$, $\beta = 1+2i$, $\mu=1+1$, we report the maximum Floquet exponent $\zeta(x)$ (blue curve) versus the semi-analytical approximations $\Sigma_0$ (green curve) and $\Sigma_1$ (red curve), for three different values of $a$ and fixed $b=0$: $a=3$ (left panel), $a=4$ (middle panel) and $a=5$ (right panel).}
   \label{fig:Floapp}
\end{figure}

By using these ideas, we can, in principle, compute for all $k$ the functions $\Sigma_k(\Lambda)$, allowing to infer about the stability of the reference limit cycle solution. In Fig.~\ref{fig:Floapp}, we report the maximum Floquet exponent $\zeta(x)$ (blue curve) versus the semi-analytical approximations $\Sigma_0$ (green curve) and $\Sigma_1$ (red curve), for the choice of parameters $\sigma = 1$, $\beta = 1+2i$, $\mu=1+1$, $b=0$ and $a=3$ (left panel), $a=4$ (middle panel) and $a=5$ (right panel). We can observe that the agreement is quite good, especially for $\Sigma_0$ that is able to reproduce the first ``bump'' of the curve $\zeta(x)$. $\Sigma_1$ allows to recover the peak of the second ``bump'' and, in particular, if it is positive or negative. Those results seem to suggest that each ``bump'' of $\zeta(x)$ corresponds to a different $k$ in Eq.~\eqref{eq:thetaanss}.

\section{Conclusions and Perspectives}
\label{sec:conc}
In this work we have studied the synchronization of identical Stuart-Landau oscillators nonlinearly coupled via a complex network, the latter being symmetric or directed. To the best of our knowledge, a similar problem had not yet been considered in the literature. Our results thus filled this gap, improving onto existing literature. We have been able to determine the necessary conditions for complete synchronization to emerge and we have shown that they depend on both the network structure and the model parameters, strengthening once again the interplay between topology and dynamics. Some choices of the nonlinear coupling are easier to be studied, e.g., the resonant case, because the linear stability can be analytically tackled by resorting to the dispersion relation instead of the Master Stability Function; the remaining non-resonant cases have been more challenging and required the introduction of Floquet theory. By means of the Jacobi-Anger expansion, we have been able to propose a semi-analytical approximation of the maximum Floquet exponent allowing to determine the emergence of complete synchronization in this case as well. 

We believe that the framework hereby introduced poses the basis for future extensions in the higher-order setting~\cite{battiston2020networks,natphys,bick2023higher,boccaletti2023structure,muolo2024turing,millan2025topology,battiston2025collective}. Note that, synchronization of coupled Stuart-Landau oscillators has already been studied on hypergraphs with a linear coupling however in a simplified setting~\cite{carletti2020dynamical}. Indeed, it has successively been shown that, in order to deal with "effective" higher-order interactions, the coupling needs to be nonlinear~\cite{neuhauser2020multibody} and non-additive~\cite{muologallo}, otherwise it can be decomposed onto pairwise interactions. Given that the Master Stability Function has been generalized to the case of identically coupled chaotic oscillators on symmetric~\cite{gambuzza2021stability} and directed~\cite{gallo2022synchronization,della2023emergence} hypergraphs, it is natural to further extend the theory hereby developed for coupled Stuart-Landau oscillators to the higher-order setting, which is something we aim to do in future works.

\noindent{\textbf{Acknowledgment}}
{The work of R.M. is supported by a JSPS postdoctoral fellowship, grant 24KF0211. H.N. acknowledges JSPS KAKENHI 25H01468, 25K03081, and 22H00516 for financial support.}


\bibliographystyle{plain}
\bibliography{biblio}

\end{document}